\pdfoutput=1
\documentclass[a4paper,11pt]{article}

\usepackage{amsmath,amssymb,amsfonts,mathrsfs}
\usepackage{dsfont}
\usepackage{tikz}
\usepackage{tikz-cd}
\usepackage{todonotes}
\usetikzlibrary{cd}
\usepackage[all]{xy}
\usetikzlibrary{shapes,arrows}
\usepackage{mathtools}

\usepackage{jheppub}

\def\bZ{\mathbb{Z}}
\def\bR{\mathbb{R}}
\def\bQ{\mathbb{Q}}
\def\bC{\mathbb{C}}
\def\bP{\mathbb{P}}

\def\bF{\mathbb{F}}

\def\cC{\mathcal{C}}
\def\cH{\mathcal{H}}
\def\cL{\mathcal{L}}

\def\cN{\mathcal{N}}
\def\cM{\mathcal{M}}

\def\cC{\mathcal{C}}
\def\cQ{\mathcal{Q}}
\def\cT{\mathcal{T}}
\def\cZ{\mathcal{Z}}

\def\MCG{\mathrm{MCG}}
\def\Aut{\text{Aut}}
\def\Arf{\text{Arf}}

\def\bp{\begin{pmatrix}}
\def\ep{\end{pmatrix}}

\def\sD{{\mathscr{D}}}

\newcommand{\Tr}{{\rm Tr}}

\def \PD {\mathop{\mathrm{PD}}}
\def \mod {\mathop{\mathrm{mod}}}

\makeatletter
\newcommand\longleftrightarrowfill@{%
  \arrowfill@\leftarrow\relbar\rightarrow}
\makeatother

\numberwithin{equation}{section} 

\title{Modularity of Vafa-Witten Partition Functions from SymTFT}

\author[a,d]{Jin Chen,}
\author[b,c]{Wei Cui,}
\author[c,b]{Babak Haghighat,}
\author[c]{Youran Sun}

\affiliation[a]{Department of Physics, Xiamen University, Xiamen, 361005, China}
\affiliation[b]{Beijing Institute of Mathematical Sciences and Applications (BIMSA), Huairou District, Beijing 101408, P. R. China}

\affiliation[c]{Yau Mathematical Sciences Center, Tsinghua University, Beijing, 100084, China}

\affiliation[d]{Peng Huanwu Center for Fundamental Theory, Hefei, Anhui 230026, China}

\emailAdd{zenofox@gmail.com}
\emailAdd{cwei@bimsa.cn}
\emailAdd{babakhaghighat@tsinghua.edu.cn}
\emailAdd{syouran0508@gmail.com}

\preprint{}

\abstract{
The 6d (2,0) theory of $N$ M5 branes compactified on the product geometry $T^2\times S$, where $S$ is a K\"ahler 4-manifold, can be studied in two different limits. In one limit, the size of $T^2$ is taken to zero and together with a topological twist one arrives at the Vafa-Witten partition function on $S$. On the other hand, taking the size of $S$ to zero leads to a 2d $\mathcal{N}=(0,4)$ theory. This gives rise to a 2d-4d correspondence where the Vafa-Witten partition functions are identified with the characters of the 2d theory. In this paper, we test this conjecture for Hirzebruch and Del Pezzo surfaces by employing the technique of SymTFT to show that the modular transformation properties of the two sides match. Moreover, we construct modular invariant 2d absolute partition functions and verify that they are invariant under gauging of a discrete symmetry at the self-dual point in coupling space. This provides further hints for the presence of duality defects in the 2d SCFT.
}

\addtocontents{toc}{\protect\setcounter{tocdepth}{2}}

\begin{document}
\hspace{\fill} USTC-ICTS/PCFT-24-33\\

\maketitle
\flushbottom


\section{Introduction}

Compactification of 6d (2,0) theories on a 2-torus leads to $\mathcal{N}=4$ Super-Yang-Mills (SYM) theory in four dimensions.
Such theories are conjectured to admit Montonen-Olive duality \cite{Montonen:1977sn}, also known as S-duality, where electric charges at weak coupling become equivalent to magnetic charges at strong coupling. From a six-dimensional perspective, S-duality arises from the action of $SL(2,\mathbb{Z})$ on the complex structure $\tau$ of the torus which corresponds to the complexified gauge coupling
\begin{equation}
    \tau = \frac{\theta}{2\pi} + \frac{4\pi i}{g^2},
\end{equation}
where $\theta$ is the theta angle of Yang-Mills theory and $g$ is the real gauge coupling parameter. A first non-trivial test of this duality in the pure gauge theoretic setting was provided by Sen \cite{Sen:1994yi} who showed the existence of certain two-monopole bound states with dyonic charge. The seminal paper of Vafa and Witten \cite{Vafa:1994tf} then provided further support by computing instanton partition functions of topologically twisted $\mathcal{N}=4$ SYM on four-manifolds and showing that they transform as vector-valued modular forms under the action of $SL(2,\mathbb{Z})$. It was already remarked in \cite{Vafa:1994tf} that for K\"ahler manifolds with $b_2^+=1$ the relevant partition functions are not exactly holomorphic in $\tau$ but rather belong to the class of so-called Mock Modular Forms (see for example \cite{Dabholkar:2012nd} for an introduction). The nonholmorphicity amounts to the fact that the relevant partition functions suffer from a holomorphic anomaly which can be traced back to the non-compactness of the Coulomb branch on such manifolds \cite{Dabholkar:2020fde}. Exact modular invariance fails from yet another perspective, as the corresponding partition functions are vector-valued modular forms. This can be traced back to the characterization of 6d (2,0) theories as relative quantum field theories where there is not a unique partition function but rather a vector \cite{Witten:1998wy, Witten:2009at}.
Thus putting the theory on the geometry $T^2 \times S$ and scaling down the size of $T^2$ gives rise in a vector-valued partition function of the resulting conformal theory on $S$. By now, many such partition functions have been obtained, using various methods, for $S$ being a Del Pezzo or Hirzebruch surface 
\cite{Vafa:1994tf,Alim:2010cf,Bringmann:2010sd,Manschot:2010nc, Manschot:2011ym, Manschot:2011dj, Haghighat:2012bm,Manschot:2014cca,Manschot:2017,Gottsche:2017vxs,Bringmann:2018cov, Alexandrov:2019rth, Alexandrov:2020bwg, Alexandrov:2020dyy, Manschot:2021qqe,Chattopadhyaya:2023aua}. 
One can then form linear combinations of partition functions which are invariant under the internal symmetries of the four-manifold $S$, see for example \cite{Manschot:2021qqe}, and thus arrive at absolute $\mathcal{N}=4$ theories as classified in \cite{Bashmakov:2022jtl}.

Turning the story around, we can also consider the limit where the size of $S$ goes to zero, resulting in a dual 2d description.
The resulting theory is a (conformal) sigma model with $\mathcal{N} = (0,4)$ supersymmetry \cite{Alim:2010cf,Haghighat:2011xx,Haghighat:2012bm,Dabholkar:2020fde}. 
In cases where $S$ is a Hirzebruch or Del Pezzo surface, the target space of the sigma model is the moduli space of magnetic monopoles and thus non-compact \cite{Haghighat:2011xx,Haghighat:2012bm}. 
A similar argument can be made in the case of $\mathbb{P}^2$ \cite{Dabholkar:2020fde}. 
     
This results in a holomorphic anomaly of the corresponding elliptic genera which in fact can be identified with the Vafa-Witten partition functions in the dual picture \cite{Alexandrov:2020bwg, Alexandrov:2020dyy}. However, these 2d partition functions still correspond to a relative theory as they transform as vectors under S-duality.
Given recent progress in the construction of absolute theories in 2d using the machinery of generalized symmetry 
\cite{Verlinde:1988sn, Bhardwaj:2017xup,Chang:2018iay,
Komargodski:2020mxz,
Hayashi:2022fkw,Roumpedakis:2022aik,Kaidi:2022uux,Choi:2022jqy,Cordova:2022ieu,Antinucci:2022eat,Bashmakov:2022jtl,Damia:2022rxw,Damia:2022bcd,Choi:2022rfe,Lu:2022ver,Bhardwaj:2022lsg,Lin:2022xod,Apruzzi:2022rei,GarciaEtxebarria:2022vzq, Benini:2022hzx, Wang:2021vki, Chen:2021xuc, DelZotto:2022ras,Bhardwaj:2022dyt,Brennan:2022tyl,Delmastro:2022pfo, Heckman:2022muc,Freed:2022qnc,Freed:2022iao,Niro:2022ctq,Mekareeya:2022spm,Antinucci:2022vyk,Chen:2022cyw,Karasik:2022kkq,Cordova:2022fhg,Decoppet:2022dnz,GarciaEtxebarria:2022jky,Choi:2022fgx,Yokokura:2022alv,Bhardwaj:2022kot,Bhardwaj:2022maz, Hsin:2022heo,Heckman:2022xgu,Antinucci:2022cdi,Apte:2022xtu, Chang:2022hud, Garcia-Valdecasas:2023mis, Delcamp:2023kew, Bhardwaj:2023zix, Yu:2023nyn,Lawrie:2023tdz,Santilli:2024dyz,Perez-Lona:2023djo,Arbalestrier:2024oqg,Copetti:2024rqj,Li:2023knf,Braeger:2024jcj,Baume:2023kkf,Heckman:2024oot,Heckman:2024obe,Apruzzi:2024htg,Bhardwaj:2024kvy,Bhardwaj:2024wlr,Bhardwaj:2024qrf,Bhardwaj:2023bbf,Bhardwaj:2023fca,Bhardwaj:2023idu,Apruzzi:2023uma,Diatlyk:2023fwf,Damia:2023ses,Antinucci:2024zjp,Argurio:2024oym,Liu:2024znj,Ambrosino:2024ggh,Okada:2024qmk,Kaidi:2024wio,Bonetti:2024cjk,DelZotto:2024tae,Hasan:2024aow,Cordova:2024vsq,Antinucci:2023ezl,Copetti:2023mcq,Cordova:2023her,Cordova:2023jip,Cordova:2023qei,Cordova:2024ypu,Sun:2023xxv,Nardoni:2024sos,Brennan:2024fgj,Cordova:2023bja,Choi:2023vgk,Seifnashri:2024dsd,Seiberg:2024gek,Choi:2023pdp,Choi:2023xjw,Choi:2024rjm, Cui:2024cav}
and SymTFTs \cite{Apruzzi:2021nmk, Apruzzi:2022dlm,Freed:2022qnc,vanBeest:2022fss,Kaidi:2022cpf,Kaidi:2023maf},
the goal of the current paper is to identify such absolute 2d partition functions for the 6d (2,0) theory of $N$ M5 branes compactified on $S$. The corresponding absolute theories were already classified, for a range of four-manifolds, in \cite{Chen:2023qnv, Bashmakov:2023kwo}. We now aim at finding the corresponding chiral partition functions. To this end, the strategy will be as follows. We obtain a 3d SymTFT by reducing the 7d SymTFT of the (2,0) theory on $S$, and subsequently put this 3d topological theory on a slab with a dynamical boundary condition at one end and a topological one at the other. This allows us to identify absolute 2d theories with modular invariant partition functions. Moreover, when the four-manifold is a Hirzebruch surface $\bF_l$ with $l=0 \mod 2$, we show how the corresponding 2d mock modular partition functions are invariant under gauging of a discrete global symmetry at a particular self-dual point in coupling space. One consequence of this is the presence of duality defect lines.

The organization of the paper is as follows. In Section \ref{sec:6dcomp} we identify SymTFTs arising from compactification of the 6d (2,0) SCFT of $N$ M5 branes on K\"ahler four-manifolds. We construct the transformation properties of the corresponding boundary partition functions under the modular group from the transformation properties of absolute 2d CFTs with $\bZ_N$-symmetry. We then proceed with a review of Vafa-Witten partition functions and subsequently derive their modular transformation properties from the SymTFT and the reduction of the 6d anomaly polynomial. In Section \ref{sec:Hirze} we then specify to Hirzebruch surfaces $\bF_l$ and show that there exists a self-dual point in coupling space where the corresponding 2d mock modular partition functions are invariant under $\bZ_N$-gauging for $l=0 \mod 2$. In Section \ref{sec:delPezzo} we specify to del Pezzo surfaces and analyze the properties of the corresponding 2d partition functions.

\section{Compactification of 6d SCFTs on four-manifolds}
\label{sec:6dcomp}

6d SCFTs on a closed six-dimensional manifold $M_6$ are in general relative theories. They can be understood as living on the boundary of a 7d topological field theory with $\partial W_7 = M_6$. The partition function of 6d SCFTs is not a number, but a vector in the Hilbert space of the 7d bulk theory. 
We will focus on the 6d $\cN=(2,0)$ SCFT of type $A_{N-1}$. It is a relative theory with defect group $\bZ_N$. The 7d bulk TQFT associated with this theory is described by the action \cite{Witten:1998wy}
\begin{equation} \label{eqn:symTFT-7d}
    S_{7d} = \frac{N}{4\pi} \int_{W_7} c \wedge dc~,
\end{equation}
where $c \in H^3(W_7, U(1))$ is a 3-form field. There are discrete 3-form fluxes $\Phi(\xi)$, valued in the defect group $\bZ_N$, given by 
\begin{equation}
    \Phi(\xi) = \exp{\left(\frac{2\pi i}{N} \int_{\PD(\xi)} c \right)},
\end{equation}
with $\xi \in H^3(M_6,\bZ_N)$. They satisfy a Heisenberg algebra
\cite{Witten:1998wy, Tachikawa:2013hya}  
\begin{equation}
    \Phi(\xi) \Phi(\xi') = e^{{2 \pi i \over N} \langle \xi,\xi' \rangle}\Phi(\xi')\Phi(\xi)~,\quad \xi,\xi'\in H^3(M_6,\bZ_N),
\end{equation}
where $\langle \xi,\xi' \rangle$ is the intersection pairing of $H^3(M_6,\bZ_N)$.

The partition vector of the 6d SCFT can be understood as a state in the Hilbert space $\cH(M_6)$ of the 7d TQFT. 
In order to specify this state, one needs to specify a maximal isotropic sublattice or polarization $\cL$ of $H^3(M_6, \bZ_N)$ such that 
\begin{equation}
    \langle \xi, \xi'\rangle = 0,\qquad \forall \; \xi,\xi' \in \cL.
\end{equation}
%
Each choice of a maximally isotropic sublattice $\cL$ of $H^3(M_6, \bZ_N)$ gives a vector $|\cL,0 \rangle$ in the Hilbert space $\cH(M_6)$ satisfying 
\begin{equation}
\Phi(\xi) |\cL,0 \rangle = |\cL ,0 \rangle, \qquad\quad \forall \; \xi \in \cL\,.
\end{equation}
The other states are obtained by acting with
\begin{equation}
\Phi(\xi')  |\cL ,0 \rangle = |\cL , \xi' \rangle, \qquad \quad  \forall \, \xi' \in \cL^\perp,
\end{equation}
where $\cL^\perp = H^3(M_6,\mathbb Z_N)/\cL$ is the Pontryagin dual of $\cL$. 
Given a choice of polarization $\cL$, the ``partition vector'' of the 6d SCFT is given by \cite{Bashmakov:2022uek}
\begin{equation}
| A_{N-1} \rangle = \sum_{\xi' \in \cL^\perp} \cZ_{\cL,\xi'}[M_6] |\cL, \xi'\rangle~,
\end{equation}
where the coefficients are the conformal blocks \cite{Witten:2009at} given by 
\begin{equation}
    \cZ_{\cL,\xi'}[M_6] = \langle \cL, \xi'| A_{N-1} \rangle.
\end{equation}
Once the polarization is fixed, we will suppress the $\cL$ dependence and write them as $\cZ_{\xi'}[M_6]$ in the following.

\subsection{Symmetry TFT}

Consider 6d SCFTs on $M_6=S \times T^2$ where $S$ is a simply connected, closed 4-manifold with $b_1=0$. The intersection form of $S$ is defined as \footnote{The intersection form is only defined on the torsion-free part of $H^2(S,\bZ)$.}
\begin{equation} \label{eq:QM4}
\begin{aligned}
     Q:  H^2(S,\bZ) \times H^2(S,\bZ) \;  &\to \; \bZ.
\end{aligned}
\end{equation}
%
For a given basis $\{e_i\}_{i=1,2,\ldots, b_2}$ of $H^2(S,\bZ)$, the intersection form can be represented by a symmetric, unimodular matrix 
\begin{equation} \label{eq:Qij}
    Q_{ij} =  \int_{S} e_i \wedge e_j, \quad i,j = 1,2,\ldots, b_2.
\end{equation}
The number of positive and negative eigenvalues of $Q$ are denoted by $b_2^+$ and $b_2^-$. The Euler characteristic and signature are
\begin{equation}
    \chi = 2+ b_2^+ + b_2^-,\qquad \sigma = b_2^+ - b_2^-.
\end{equation}
These are the two basic topological invariants that will be used later. 
For a pair of cycles $\mu,\nu \in H^2(S,\bZ)$, the intersection pairing $(-,-)$ is defined to be  
\begin{equation} \label{eq:pairing1}
     (\mu,\nu) = \int_{S} \mu \wedge \nu =\mu^i Q_{ij} \nu^j \in  \bZ,
\end{equation}
where $\mu^i$ and $ \nu^j$ are components of $\mu$ and $\nu$ with $i,j=1,2,\ldots, b_2$. 

We will focus on cases where $S$ is a K\"ahler 4-manifold. Let $K_S$ be the canonical class given by the first Chern class as $K_S=-c_1(S)$. 
It is characteristic in the sense that for $\forall \lambda \in H^2(S,\bZ)$, one has   
\begin{equation} \label{eq:m4fact1}
    (K_S,\lambda ) = (\lambda,\lambda),\quad \mod\; 2 .
\end{equation}
When $S$ has almost complex structure, it also satisfies \cite{Manschot:2021qqe}
\begin{equation} \label{eq:m4fact2}
    (K_S,K_S) = \sigma, \quad \mod\; 8.
\end{equation}
The canonical class is the integer lift of the second Stiefel-Whitney class $w_2(S)$ in $H^2(S,\bZ)$. If $S$ is spin, the canonical class $K_S$ and the intersection form $Q$ in \eqref{eq:QM4} are even.

Most of 4-manifolds are non-spin with $w_2 \neq 0$. One can define the ${\rm Spin}^c$ structure on $S$ as $TS \otimes K^{-1/2}_S$. Then the ${\rm Spin}^c$ field strength is specified by taking  
\begin{equation}
    \tilde \mu = \mu - \frac{K_S}{2} \in H^2(S,\bZ).
\end{equation}
Since $\mu$ and $\tilde \mu$ are $\bZ_N$ valued, one can write the above as 
\begin{equation}\label{eq:defmutilde}
    \tilde \mu = \mu - \frac{N w_2}{2} \in H^2(S,\bZ_N).
\end{equation}
where $w_2$ is the second Stiefel–Whitney class satisfying $w_2 = K_S\;\mod \;2$. So, $w_2/2 = K_S/2\;\mod \;1$ and $Nw_2/2 = K_S/2\;\mod \;N$.



\begin{figure}
    \centering
    \begin{tikzpicture}[scale=0.75]
        \fill[blue!20] (0.00,0.00) rectangle (6,3);
        \fill[blue!20] (13,0) rectangle (19,3);
        \draw[ thick] (13,0) -- (13,3);
        \draw[ thick] (19,0) -- (19,3);
	\draw[thick] (6,0) -- (6,3);
        \draw[thick,  ->] (6.7,1.5) -- (12.3,1.5);
        \node[above] at (9,1.6) {Shrink $S$};
	\node at (16,1.5) {SymTFT($\cC$)};
 \node at (3,1.5) {7d TQFT};
 \node[below] at (6,0) {$|A_{N-1}\rangle $}; 
 \node[below] at (13,0) {$\langle L(B)|$};
 \node[below] at (19,0) {$|\cT_N[S]\rangle $}; 
 \node[above] at (19,3) {$x=\epsilon$};
 \node[above] at (13,3) {$x=0$};
	\end{tikzpicture}
    \caption{Compactification of the 7d/6d coupled system on $S$ with maximal isotropic sublattice $L$ leads to a 2d theory $T_{N}[S]$ on $T^2$ and its SymTFT on $T^2 \times I_{(0,\epsilon)}$ with topological boundary condition $\langle L(B)|$.}
    \label{Fig:compactM4}
\end{figure}
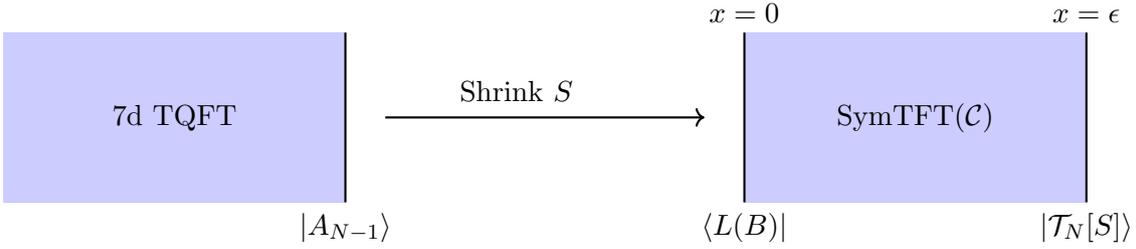

Consider the compactification of the 6d SCFT on $S$. Under the VW twist, which will be introduced later, the low-energy effective theory is a 2d $\cN=(0,4)$ theory denoted by $T_N[S]$ \cite{Maldacena:1997de}. Without the specification of the polarization in $H^2(S,\bZ_N)$, it is a relative theory. Similar to the 6d SCFT, $T_N[S]$ can be understood as living on the boundary of a three-dimensional symmetry TFT (SymTFT). The 2d theory $T_N[S]$ is a relative theory living on the dynamical boundary of the 3d SymTFT. 
To obtain absolute theories on $T^2$, one needs to specify a maximal isotropic sublattice $L \subset H^2(S,\bZ_N)$ such that 
\begin{equation} \label{Eqn:TopBndCon}
    ( \mu, \mu') = 0 ~, \quad \forall \; \mu,\mu'\in L~.
\end{equation}
As discussed in \cite{Aharony:2013hda, GarciaEtxebarria:2019caf}, 
different choices of $L$ lead to different absolute theories of $T_N[S]$ with different global properties. Equivalently, the choice of maximal isotropic sublattices $L$ corresponds to the the topological boundary condition $| L,0\rangle$ of the SymTFT \cite{Gukov:2020btk}. 
The partition function of the absolute theories are given by 
\begin{equation}
    Z_{L}[T^2] = \langle L,0| \cT_N[S]\rangle\;
\end{equation}
which are different linear combinations of the partition vectors.

To specify the global variants, one also needs to choose a specific representative of the non-trivial classes of $L^\perp \otimes H^1(T^2, \bZ_N)$, with $L^\perp= H^2(S, \bZ_N)/L$. The
choice of representatives in $L^\perp$ determines if the 2d theory is stacked with a possible SPT phase while the choice of elements in $H^1(T^2,\bZ_N)$  determines the background fields for
the corresponding zero-form symmetries. The partition functions of different global variants are denoted by \cite{Bashmakov:2022jtl}
\begin{equation}
    Z_{L,B}[T^2] = \langle L,B| \cT_N[S]\rangle\;, \quad B\in L^{\perp}.
\end{equation}
We will identify them with the VW partition functions using 4d/2d correspondence.   

The cohomology lattice on $S \times T^2$ splits via the K{\"u}nneth formula as 
\begin{equation*}
H^3(M_6, \bZ_N) \cong   H^2(S,\bZ_N) \otimes H^1(T^2, \bZ_N)~. 
\end{equation*}
Given two three-forms $\xi= \mu\wedge \eta$ and $\xi'= \mu' \wedge \eta'$ in $H^3(M_6, \bZ_N)$ with $\mu,\mu' \in H^2(S,\bZ_N)$ and $\eta,\eta' \in H^1(T^2,\bZ_N)$, their intersection pairing decomposes accordingly as
\begin{equation} \label{Eqn:decomPairing}
    \langle \xi,\xi' \rangle = (\mu,\mu') \times \langle \eta,\eta' \rangle\,,
\end{equation}
where $\langle-,-\rangle$ is the standard anti-symmetric intersection pairing on $T^2$ while $(-,-)$ is the intersection paring of $S$ defined above.

The SymTFT can be determined from the dimensional reduction of the 7d action in equation \eqref{eqn:symTFT-7d}. Expanding the three-form $c=\sum_{i=1}^{b_2} a^i \wedge e_i$ where $a^i \in H^1(T^2,\bZ_N)$ and $\{e_i\}_{i=1,2,\ldots, b_2}$ is a basis of $H^2(S,\bZ)$, 
and compactifying the 7d action \eqref{eqn:symTFT-7d} leads to the following action in 3d
\begin{equation} \label{eq:A3d}
    S_{3d} = \frac{N}{4\pi} \sum_{i,j} Q_{ij} \int_{W_3}  a^i \wedge d a^j.
\end{equation}
This is an abelian Chern-Simons theory with level matrix $K_{ij} \equiv N Q_{ij}$. Let us define the lattice $\Lambda =  \bZ\langle e_1,\ldots,e_{b_2}\rangle = H^2(S,\bZ)$ with bilinear form $\left(e_i,e_j\right)_K = K_{ij}$. The dual lattice $\Lambda^*$ is generated by vectors $e^{*i}$ defined by $e^{*j}(e_i) = \delta^j_{~i}$. It is easy to see that $e^{*i} = \left(K^{-1}\right)^{ij} e_j$ as
\begin{equation}
    e^{*j}(e_i) \equiv \left(K^{-1}\right)^{jk}\left(e_k,e_i\right)_K = \left(K^{-1}\right)^{jk} K_{ki} = \delta^j_{~i},
\end{equation}
and hence $\Lambda^* = H^2(S,1/N \bZ)$.
If $K$ defines an even lattice, which happens when $S$ is a spin manifold or $N$ is even, the SymTFT is bosonic. Otherwise, it is a spin Chern-Simons theory \cite{Belov:2005ze}. The discriminant group is
\begin{equation}
    \mathscr{D} = \Lambda^*/\Lambda = \bZ^{b_2}/(NQ\bZ^{b_2}) = (\bZ_N)^{b_2}.
\end{equation}

The anyons arise from the reduction of the discrete three-form flux on $S$ in the following way. Take the three-form flux to be $\Phi(\eta \times \tilde \mu)$ with $\eta \in H^1(T^2,\bZ_N)$ and $\tilde \mu $ defined in Eq \eqref{eq:defmutilde}. 
The anyons after the reduction are given by 
%

%
\begin{equation}
\phi_{\tilde \mu^*}(\eta)=
\exp{\left(\frac{ 2\pi i }{N}\int_{\PD(\eta)} a^i Q_{ij} \tilde \mu^j \right)} = \exp{\left(\frac{2\pi i}{N} \int_{\PD(\eta)} a^i  (\tilde \mu^*)_i \right)},
\end{equation}
where 
\begin{equation} \label{eq:shiftedmudual}
      (\tilde \mu^*)_i= Q_{ij} \tilde \mu^j  = Q_{ij} \left( \mu^j - \frac{N(w_2)^j}{2}\right).
\end{equation}
The topological spin of the anyon with charge $\tilde \mu^*$ is a non-degenerate homogeneous quadratic function $\theta: \sD \to \bQ/\bZ$ defined by \cite{Belov:2005ze}
\begin{equation} 
    \theta(\tilde \mu^*) \equiv  \exp{\left( \pi i \tilde \mu^*_i  \left(K^{-1}\right)^{ij} \tilde \mu^*_j  \right)} 
    = \exp{\left( 2\pi i \frac{(\tilde \mu,\tilde \mu)}{2N}  \right)}
    = \exp{\left( \frac{2\pi i}{N} q_{w_2}(\mu)   \right)},
\end{equation}
where 
\begin{equation} \label{eq:defq}
    q_{w_2}(\mu) = \frac{1}{2} (\mu-\frac{Nw_2}{2}, \mu-\frac{Nw_2}{2}),\quad \mod \; N.
\end{equation}
The braiding between two anyons with charge $\tilde \mu^*$ and $\tilde \nu^*$ is 
\begin{equation} 
    B( \tilde \mu^*, \tilde \nu^*) \equiv \frac{\theta(\tilde \mu^*+\tilde \nu^*)\theta(0)}{\theta(\tilde \mu^*)\theta(\tilde \nu^*)}=
   \exp\left(\frac{2\pi i}{N} (\mu,\nu) \right).
\end{equation}
The S-matrix and T-matrix of the corresponding TQFT are given by
\begin{equation} \label{eq:STsymTFT}
\begin{aligned}
    &S(\tilde \mu^*,\tilde \nu^*) = \frac{1}{\sqrt{|\sD|}} \exp\left(\frac{2\pi i}{N} (\mu,\nu) \right),
    \\
    &T(\tilde \mu^*,\tilde \nu^*)  = \exp{\left(\frac{2 \pi i}{N} q_{w_2}(\mu)- \frac{2 \pi i c}{24} \right)}  \delta_{\mu \nu}, 
\end{aligned}
\end{equation}
where $c$ is the chiral central charge, satisfying the Gauss sum constraint 
\begin{equation}
    \sum_{\tilde \mu^* \in \sD} \theta(\tilde \mu^*) = \sqrt{|\sD|} e^{2\pi i c/8}.
\end{equation}

The partition vector of the compactified 6d SCFT is a state in the Hilbert space of the 3d theory which is given by 
\begin{equation}
    |T_N[S] \rangle = \sum_{\mu \in \cL^\perp} Z_{\cL,\mu}[T^2] |\cL, \mu\rangle~,
\end{equation}
where $\cL$ is a polarization of $H^3(M_6,\bZ_N)$ independent of $S$, $\mu \in H^2(S,\bZ_N) \subset \cL^{\perp}$ and the coefficients $Z_{\cL,\mu}[T^2]$ are 2d conformal blocks which equivalently can be understood as partition functions with the anyon insertion $\phi_{\tilde{\mu}^*}$.

The $S/T$-matrices can also be established from the boundary CFT perspective. In the case of $N$ M5 branes wrapping $\mathbb F_0$, for simplicity, the resulting SymTFT is the ordinary $\mathbb Z_N$ Witten-Dijkgraaf theory with the Lagrangian
\begin{align}
        S_{3d}=\frac{N}{2\pi}\int \hat a\wedge da.
\end{align}
The 3d partition function on a solid torus $D_\epsilon\times S^1$ can be evaluated with insertion of an anyon $L_{\vec a}$, say $Z_{3d}[\epsilon; L_{\vec a}]$, where $\epsilon$ is the distance from the anyon line to the rim of the disk, as shown in Figure \ref{fig:BCFT}.

\begin{figure}[ht]
\centering
\begin{tikzpicture}[scale=0.75]
    \draw[dotted] (0,0)ellipse[x radius=2,y radius=1];
    \draw (-2,0)--(-2,4);
    \draw (2,0)--(2,4);
    \draw (0,4)ellipse[x radius=2,y radius=1];
    \fill[blue!20] (0,4)ellipse[x radius=1.99,y radius=0.99];
    \draw (2,0) arc (0:-180:2 and 1);
    \path[draw,red] (0,0)--(0,2.95);
    \path[draw] (0,0)--(2,0);
    \draw(1, .2)node{$\epsilon$};
    \path[draw,red,dotted] (0,3.05)--(0,4);
    \fill[red] (0,0) circle (1.5pt);
    \fill[red] (0,4) circle (1.5pt);
    \draw(0,2)node[right]{$L_{(m,n)}$}; 
    \draw(0, 5.4)node{$D_\epsilon$}; 
    \draw(-2.4, 2)node{$S^1$};
    \draw[thick, ->] (3,1) -- (7,1);
    \node[above] at (5,1.2) {$\epsilon \to 0$};
    \draw[dotted] (10,0)ellipse[x radius=2,y radius=1];
    \draw (8,0)--(8,4);
    \draw (12,0)--(12,4);
    \draw (10,4)ellipse[x radius=2,y radius=1];
    \fill[blue!20] (0,4)ellipse[x radius=1.99,y radius=0.99];
    \draw (12,0) arc (0:-180:2 and 1);
    \path[draw,red] (10,-1)--(10,3);
    \fill[red] (10,-1) circle (1.5pt);
    \fill[red] (10,3) circle (1.5pt);
    \draw(10,2)node[right]{$\mathcal L_n$}; 
    \draw(10,-1.4)node{$m$-th state}; 
    \draw(10,5.4)node{$S^1$}; 
    \draw(7.6,2)node{$S^1$};
    \path[draw,red,dotted] (0,3.05)--(0,4);
    \fill[red] (0,4) circle (1.5pt);
\end{tikzpicture}
\caption{}
\label{fig:BCFT}
\end{figure}

When the $\epsilon$-slab in the solid torus goes to zero, it connects to the partition function of the 2d boundary CFT on the torus $S^1\times S^1$ with $\mathbb Z_N$-symmetry. That is the anyons of type $L_{\vec a}=L_{(m,0)}$ condense, while the anyons of type $L_{\vec a}=L_{(0,n)}$ serve as $\mathbb Z_N$-topological defect lines, denoted by $\mathcal L_n$ for $n\in\mathbb Z_N$, on the 2d boundary CFT. 
The inserted anyon line $L_{\vec a}=L_{(m,n)}$ corresponds to the $m$-th state in the defect Hilbert space of $\mathcal H_{n}$, where $\mathcal H_n$ denotes the defect Hilbert space twisted by the symmetric defect line $\mathcal L_n$. Therefore we can establish the relation
\begin{align}
    Z_{T^2}[L_{(m,n)}]\equiv\lim_{\epsilon\rightarrow 0}Z_{3d}[\epsilon; L_{(m,n)}]=\frac{1}{N}\sum_{l\in\mathbb Z_N}\omega^{lm}Z_{T^2}^{(l,n)}\,,
\end{align}
where $\omega$ is a $N$-th primitive root of unity, $\omega=\exp{\frac{2\pi i}{N}p}$, for $\gcd(p,N)=1$, $Z_{T^2}^{(l,n)}$ is the 2d CFT partition function dressed with the $\mathbb Z_N$-symmetric lines $g^l$ horizontally and $g^n$ vertically, and thus the projector $\frac{1}{N}\sum_l\omega^{lm}$ projects onto the $m$-th state in the defect Hilbert space $\mathcal H_{g^n}$. Further using the identity
\begin{align}
    \frac{1}{N}\sum_{l\in\mathbb Z_N}\omega^{l(m-n)}=\delta_{mn}\,,
\end{align}
one can rewrite 
\begin{align}
    Z_{T^2}^{(m,n)}=\sum_{l\in\mathbb Z_N}\omega^{-ml}Z_{T^2}[L_{(l,n)}].
\end{align}
Now implementing the $S$-modular transformation on $Z_{T^2}[L_{(m,n)}]$, we have
\begin{align}
    S^{\rm top}\cdot Z_{T^2}[L_{(m,n)}]=\frac{1}{N}\sum_{l\in\mathbb Z_N}\omega^{lm}Z_{T^2}^{(-n,l)}=\frac{1}{N}\sum_{k,l\in\mathbb Z_N}\omega^{lm+kn}Z_{T^2}[L_{(k,l)}]\,.
\end{align}
Therefore, we can read off the matrix elements of $S^{\rm top}$ as
\begin{align}
    S^{\rm top}_{(m,n);\,(k,l)}=\frac{1}{N}\omega^{ml+nk}\,.
\end{align}
On the other hand, for $T$-modular transformation, we similarly have
\begin{align}
    T^{\rm top}\cdot Z_{T^2}[L_{(m,n)}]&=\frac{1}{N}\sum_{l\in\mathbb Z_N}\omega^{lm}Z_{T^2}^{(n+l,n)}\notag\\
    &=\frac{1}{N}\sum_{k,l\in\mathbb Z_N}\omega^{lm-k(n+l)}Z_{T^2}[L_{(k,n)}]\notag\\
    &=\sum_{k\in\mathbb Z_N}\omega^{-kn}\delta_{mk}Z_{T^2}[L_{(k,n)}]=\omega^{-mn}Z_{T^2}[L_{(m,n)}]\,.
\end{align}
Therefore, the $T^{\rm top}$-matrix is spelt as
\begin{align}
        T^{\rm top}_{(m,n);\,(k,l)}=\delta_{mk}\delta_{nl}\omega^{-mn}\,.
\end{align}

\subsection{Vafa-Witten Theory}

Consider the 4d $\cN=4$ supersymmetric Yang-Mills theory of gauge group $SU(N)$. 
The gauge coupling $g$ together with the $\theta$-angle defines the complex parameter $\tau = \frac{\theta}{2\pi}+\frac{4\pi i}{g^2}$. Under the S-duality $\tau \to -\tau^{-1}$, the gauge group exchanges with its Langlands dual, for example, between $SU(N)$ and $PSU(N)=SU(N)/\bZ_N$ in our case. The S-duality extends to the action of the full modular group and the partition function is expected to be a modular form under $SL(2,\bZ)$. This can be studied explicitly using the partition functions of the VW theory, i.e. the topologically twisted $N=4$ supersymmetric Yang-Mills theory. We will briefly review it in the following \cite{Vafa:1994tf}. 

The global symmetry of the 4d $N=4$ Yang-Mills theory includes the Lorentz group $SU(2)_L\times SU(2)_R$ and R-symmetry $SU(4)$. 
The supercharges $\cQ_{\alpha}^{i}$ and $ \cQ_{\dot{\alpha}}^{i}$ transform as $(\bf{2,1,\bar{4}})$ and $(\bf{1,2,4})$ under $SU(2)_L\times SU(2)_R \times SU(4)$. 
Consider the decomposition of the R-symmetry $SU(2)_A \times SU(2)_B \subset SU(4)$. The VW twist identifies the new $SU(2)_R^{\text{tw}}$ as the diagonal subgroup of $SU(2)_R \times SU(2)_B$. The other $SU(2)_A$ becomes the R-symmetry of the twisted theory. The supercharges $\cQ_{\alpha}^{i}$ and $ \cQ_{\dot{\alpha}}^{i}$ transform as  
\begin{eqnarray}
& SU(2)_L\times SU(2)_R \times SU(4) & \to \quad SU(2)_L\times SU(2)_R^{\text{tw}} \times SU(2)_A, \nonumber \\
& (\bf{2,1,\bar{4}}) + (\bf{1,2,4}) & \to \quad {\bf (2,2,2)}+{\bf (1,1,2)}+{\bf (1,3,2)}.  \nonumber
\end{eqnarray}
We find two scalar supercharges denoted by $\cQ$ and $\cQ'$. They satisfy $\cQ^2=\cQ'^2=0$ and $\{\cQ,\cQ'\}=0$. One can identify them as the BRST operator and the twisted theory is topological on $S$.



The path integral of VW theory localizes on solutions of hermitian Yang-Mills equations. Let $\cM_{\Gamma,J}$ be the moduli space of these solutions where $J \in H^2(S,\bR)$ is the K\"ahler form on $S$ and $\Gamma=(N,\mu, n)$ is the Chern characters of the $U(N)$ bundle with instanton number $n$ and t'Hooft magnetic flux $\mu=-c_1(F)\in H^2(S,\bZ_N)$. The VW partition function is a (holomorphic) generating function of the Euler characteristic of $\cM_{\Gamma,J}$ given by
\footnote{Here, we assume $(N,\mu, n)$ coprime. If it is not the case, $ h_{N,\mu,J}(\tau)$ generates the rational VW invariant of $\cM_{\Gamma,J}$.}
\begin{equation}
    h_{N,\mu,J}(\tau) = \sum_{n\geq 0} \chi(\cM_{\Gamma,J}) q^{N(\Delta_F-\frac{\chi}{24})}
\end{equation}
where $q=e^{2\pi i \tau}$ and 
\begin{equation}
    \Delta_F = \frac{1}{N}\left( n - \frac{N-1}{2N} (\mu,\mu)\right)
\end{equation}
is the Bogomolov discriminant. To relive the notation, we will suppress the $J$ dependence when there are no confusion.  
As an example, the rank one VW partition function is  
\begin{equation} \label{eq:VW1M}
    h_{1,0}(\tau) = \frac{1}{\eta(\tau)^{\chi}}
\end{equation}
for any K\"ahler 4-manifold with $b_1=0$. 
The rank $N>1$ ones are usually expressed as $h_{N,\mu}(\tau) = f_{N,\mu}(\tau)h_{1,0}(\tau)^N$ where $f_{N,\mu}(\tau)$ is a function to be determined.

It is expected that the VW partition function transforms as a vector-valued modular form under $SL(2,\bZ)$. However, on $S$ with $b_2^+=1$, $h_{N,\mu}(\tau)$ are mock modular forms. One needs to add appropriate non-holomorphic terms to make them transform as a modular form. The details are left in Appendix \ref{app1}. The VW partition function after this modular completion is denoted by $\widehat h_{N,\mu}(\tau)$. The corresponding modular transformations are then \cite{Manschot:2021qqe} \footnote{Note that there is a normalization $N$ difference with \cite{Manschot:2021qqe}. There is a factor $(-1)^{(N-1)(\chi+\sigma)/4} = 1$ in our case since $\chi+\sigma = 1 \mod\; 4$.}
\begin{equation}\label{eq:STVW}
\begin{aligned} 
\widehat h_{N,\mu}(-1/\tau)&= \frac{ (-i\tau)^{-\chi/2}}{N^{b_2/2}}
\sum_{\nu \in H^2(S,\bZ_N)} \exp{\left(-\frac{2\pi i}{N} (\mu, \nu) \right)} \,\widehat h_{N,\nu}(\tau)\\
\widehat h_{N,\mu}(\tau+1)&=\exp\left( -\pi i \frac{N-1}{N} (\mu,\mu) - 2 \pi i\frac{N\chi}{24}  \right)\,\widehat h_{N,\mu}(\tau)\;.
\end{aligned}
\end{equation}
They transform as a modular form with weight $-\chi/2$. As an example, the rank one VW partition function in equation \eqref{eq:VW1M} indeed has this modular weight.

The refined VW partition function is defined by introducing the fugacity of the $SU(2)$ R-symmetry denoted by $z$. It is given by 
\begin{equation}
    h_{N,\mu,J}(\tau,z) = \sum_{n\geq 0} \Omega(\cM_{\Gamma,w;J}) q^{N(\Delta_F-\frac{\chi}{24})},
\end{equation}
where $w=e^{2\pi i z}$ and the invariants are
\begin{equation}
    \Omega(\cM_{\Gamma,w;J}) = \frac{w^{-d}}{w-w^{-1}} \sum_{i=0}^d b_i(\cM_{\Gamma,J})w^i, \qquad d = \textrm{dim}_{\bC}(\cM_{\Gamma,J}).
\end{equation}
Here, $b_i(\cM_{\Gamma,J})$ are the Betti numbers of the moduli space $\cM_{\Gamma,J}$. 
The rank one VW partition function on $S$ is given by
\begin{equation} \label{eq:VWM1R}
    h_{1,0}(\tau,z)= \frac{i}{\theta_1(\tau,2z)\eta(\tau)^{b_2-1}}.
\end{equation}
For rank $N>1$, they are given in the form of $h_{N,\mu}(\tau,z) = g_{N,\mu}(\tau,z) h_{1,0}(\tau,z)^N$ with $g_{N,\mu}(\tau,z)$ holomorphic functions to be determined. 

Similarly, on a 4-manifold with $b_2^+(S) = 1$, the refined VW partition functions are Mock Jacobi forms. After the modular completion, they transform as a vector-valued Jacobi form with  \cite{Alexandrov:2019rth, Alexandrov:2020bwg, Alexandrov:2020dyy}
\begin{equation}
\label{eq:STRVW}
\begin{aligned} 
\widehat{h}_{N,\mu}\left(-\frac{1}{\tau},\frac{z}{\tau}\right)
&=i^{2N-1}\frac{(-i\tau)^{(\sigma-\chi)/4}}{N^{b_2/2}}  
e^{ 2\pi i \frac{ m(N) z^2}{\tau}  } 
\sum_{\nu \in H^2(S,\bZ_N)}
\exp{\left(-\frac{2\pi i}{N} (\mu, \nu) \right)} \;  \widehat{h}_{N,\nu}(\tau,z), \\
\widehat{h}_{N,\mu}(\tau+1,z)&= \exp\left( -\pi i \frac{N-1}{N} (\mu,\mu) - 2 \pi i\frac{N\chi}{24}  \right) \, \widehat{h}_{N,\mu}(\tau,z),
\end{aligned}
\end{equation}
where the index $m(N)$ is 
\begin{equation} \label{eq:indexVW}
 m(N) = -\frac{1}{6}\left[(2\chi+3\sigma)(N^3-N)+3(\chi+\sigma)N\right]\;.
\end{equation}
The rank one refined VW partition function in \eqref{eq:VWM1R} has weight $-b_2/2$ and index $m(1)=-2$ corresponding to 4-manifolds with $b_2^+=1$.

\subsection{Generalized elliptic genera of $T_N[S]$}

From the perspective of 6d SCFTs, the VW partition functions on $S$ give the partition vector of a 2d relative $\cN=(0,4)$ theory $T_N[S]$. This statement is supported by the observation that the (refined) VW partition function transforms in the same way as the partition vector under modular group $SL(2,\bZ)$ up to an overall phase. 
\begin{itemize}
    \item The overall phase of the S-transformation includes a nontrivial modular weight due to the gravitational coupling on the curved 4-manifolds $S$ \cite{Vafa:1994tf} and modular index that is supposed to give the right moving central charge $c_R$ of $T_N[S]$.
    \item 
    The overall phase of the T-transformation
    is believed to give the left-moving central charge $c_L$ of $T_N[S]$.
\end{itemize}
The central charges are determined from the dimensional reduction of the anomaly polynomial. From the modularity of the VW partition function, we can get the same $c_R$ from the index up to a $1/2$ factor while for $c_L$, we can get the same result if $S$ is spin.

The 4d $\cN=4$ supersymmetric Yang–Mills theory with Lie algebra $g=su(N)$ is believed to arise from the compactification of the 6d $\cN=(2,0)$ SCFT of type $A_{N-1}$ on $T^2$ \cite{Witten:1995zh}. 
As discussed in the beginning of this section, the 6d SCFTs are relative. To obtain an absolute 4d theory, one needs to choose a polarization of $H^1(T^2,\bZ_N)$ in the compactification. Depending on different choices of the polarizations, the  gauge group of the 4d $\cN=4$ Yang-Mills theory can be $SU(N)/\bZ_k$ with $\bZ_k$ being a subgroup of $\bZ_N$ \cite{Aharony:2013hda}.

The VW theory turns out to be the theory with gauge group $SU(N)$ corresponding to the maximal isotropic sublattice $\langle \eta_A \rangle \subset H^1(T^2,\bZ_N)$ generated by the Poincare dual of the A-cycle $\eta_A$ of the torus \cite{Tachikawa:2013hya}. 
%
%
It defines a polarization in the 6d SCFT as 
\begin{equation} \label{eq:6dLA}
     \cL_A = \langle \eta_A \rangle \otimes H^2(S,\bZ_N)\, \subset H^3(M_6,\bZ_N).
\end{equation}
The 6d conformal blocks with respect to this polarization are $ Z_{\cL_A, \mu}[T^2\times S]$ with $\mu \in \cL^{\perp}_A = H^1(T^2,\bZ_N)/\langle \eta_A \rangle  \times H^2(S,\bZ_N)$. When reducing on $T^2$, one has 
\begin{equation} \label{eq:6d/4d}
    Z_{\cL_A, \mu}[T^2\times S] = h_{N,\mu}[S],
\end{equation}
where $\mu \in H^2(S,\bZ_N)$ is the t'Hooft magnetic flux in the VW partition function.


On the other hand, reducing the 6d SCFT along $S$ with the polarization \eqref{eq:6dLA} leads to a relative 2d theory $T_N[S]$. The 6d conformal blocks are expected to be the 2d ones defined in 
\begin{equation} \label{eq:6d/2d}
    Z_{\cL_A, \mu}[T^2\times S] =  Z_{\mu^*}[T^2], 
\end{equation}
where $\mu^* \in H^2(S,\bZ_N)$ labels the 2d partition function with anyon of charge $\mu^*$. A unique partition function is obtained once we choose a topological boundary condition of the SymTFT.

\begin{figure}[h]
    \centering
    \resizebox{.5\textwidth}{!}{%
    \begin{tikzpicture}[edge from parent/.style={draw,-latex}]
        \node {$Z_{\cL_A,\mu}[S \times T^2]$} [sibling distance = 5.5cm, level distance = 4cm]
        child {node {$h_{\mu}[S]$} 
            edge from parent
                node[left]  {$T^2\to 0$}}
            child {node {$Z_{\mu^*}[T^2]$} edge from parent node [right] {$S \to 0$}};
    \end{tikzpicture}}
    \caption{From 6d perspective, VW partition function $h_{\mu}[S]$ is related to the generalized elliptic genera $Z_{\mu^*}[T^2]$ with $\mu \in \Lambda$ and $\mu^* \in \Lambda$.}
    \label{fig:2d/4d}
\end{figure}
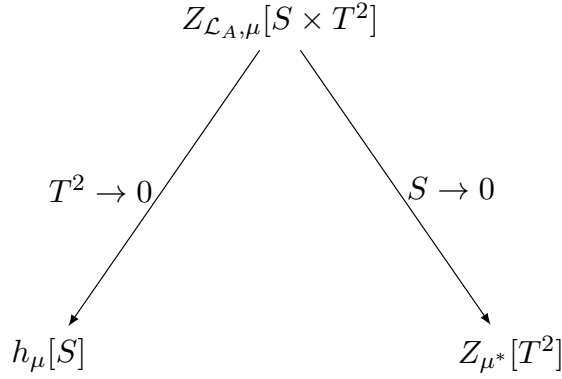

Thus, from the perspective in 6d as shown in Figure \ref{fig:2d/4d}, one can relate the VW partition function on $S$ with the 2d partition vector of $T_N[S]$ as 
\begin{equation} \label{eq:2d/4d}
    Z_{\mu^*}[T^2] = h_{N,\mu}[S].
\end{equation}

The 2d partition functions corresponding to the VW partition functions are believed to be the generalized elliptic genera defined by \cite{Cecotti:1992qh,Minahan:1998vr,Maldacena:1999bp} 
\begin{equation} \label{eq:genEG}
    Z(\tau,\bar{\tau}) = \Tr \left[(-1)^F \frac{F^2}{2} q^{L_0-c_L} \bar{q}^{\bar{L}_0-c_R}\right],
\end{equation}
where $F$ is the right-moving fermion number operator. Note that the usual elliptic genera vanishes for $\mathcal{N}=(0,4)$ theories due to an excess of fermionic zero modes. The modified index was introduced to study the BPS state counting in the MSW CFT \cite{Maldacena:1997de} of M5 branes wrapping a surface and is hence suitable for describing VW partition functions.

The central charges of $T_N[S]$ are determined from the dimensional reduction of the 6d anomaly polynomial, 
\begin{align}
\begin{aligned}
    c_L =(2\chi+3\sigma)(N^3-N)+ N\chi, \quad c_R = (2\chi+3\sigma)(N^3-N)+\frac{3N}{2}(\chi+\sigma), 
\end{aligned}
\end{align}
with chiral central charge $c_R-c_L = N(\chi+3\sigma)/2$.
The details are given in Appendix \ref{app:anomaly}.

\subsubsection*{S-transformation:}

The refined VW partition function transforms under S-duality as a vector-valued modular form as specified in equation \eqref{eq:STRVW}. One can substitute $\nu \to -\nu$ and use the symmetry of the VW partition function to replace $\widehat h_{N,\mu}$ with $\widehat h_{N,-\mu}$. The equation \eqref{eq:STRVW} becomes 
\begin{equation}
\begin{aligned}
\widehat{h}_{N,\mu}\left(-\frac{1}{\tau},\frac{z}{\tau}\right)
& =i^{2N-1}\frac{(-i\tau)^{(\sigma-\chi)/4}}{N^{b_2/2}}  
e^{ 2\pi i \frac{ m(N) z^2}{\tau}  } \sum_{\nu \in H^2(S,\bZ_N)}
\exp{\left(\frac{2\pi i}{N} (\mu, -\nu) \right)} \;  \widehat{h}_{N,-\nu}(\tau,z) \\
=&i^{2N-1}\frac{(-i\tau)^{(\sigma-\chi)/4}}{N^{b_2/2}}  
e^{ 2\pi i \frac{ m(N) z^2}{\tau}  } \sum_{\nu \in H^2(S,\bZ_N)}
\exp{\left(\frac{2\pi i}{N} (\mu, \nu) \right)} \;  \widehat{h}_{N,\nu}(\tau,z).
\end{aligned}
\end{equation}
We can see that up to an overall factor, it is the same as the S-matrix of the SymTFT which was given in \eqref{eq:STsymTFT}.

The non-trivial weight is believed to arise from the coupling with the $R^2$ terms for curved 4-manifolds $S$ \cite{Vafa:1994tf}. The index given in \eqref{eq:indexVW} gives the t'Hooft anomaly of the R-symmetry. From the dimensional reduction of anomaly polynomial \eqref{eqn:c20}, we find the t'Hooft anomaly of the R-symmetry is 
\begin{equation}
    A_R = I_4|_{c_1(R^2)} =  \frac{1}{6}\left[ (N^3-N) (2\chi+3\sigma)+\frac{3N}{2}(\chi+\sigma)\right].
\end{equation}
Comparing with the index result \eqref{eq:indexVW}, one finds that terms proportional to $(N^3-N)$, which reflect the interactions between coincident M5 branes, exactly match with the anomaly polynomial result. However, the terms from the contribution to free $N$ M5 branes are different by a factor of $1/2$. It would be interesting to obtain a better understanding of this mismatch in the future.

\subsubsection*{T-transformation:}

The T-transformation of the (refined) VW partition function is given by $\tau \to \tau+1$ because they are holomorphic in $\tau$. This does not change after the modular completion as one just adds terms containing $\im\tau$ which are invariant under the shift. 
The (refined) VW partition function under $\tau \to \tau+1$ is given in \eqref{eq:STRVW} which we recall here for convenience,
{\small
\begin{equation}
\begin{aligned}
\widehat h_{N,\mu}(\tau+1)
& = \exp\left( -\pi i \frac{N-1}{N} (\mu,\mu) - 2 \pi i\frac{N\chi}{24}  \right)\,\widehat h_{N,\mu}(\tau) .
\end{aligned}
\end{equation}}
We can see that up to an overall factor, it is the same as the T-matrix of the SymTFT in \eqref{eq:STsymTFT}:
{\small
\begin{equation}
\begin{aligned}
\widehat h_{N,\mu}(\tau+1)
&=\exp\left( -\pi i \frac{N-1}{N} (\mu,\mu) - 2 \pi i\frac{N\chi}{24}  \right)\,\widehat h_{N,\mu}(\tau) \\
&=  \exp\left( \frac{2 \pi i }{2N} (\mu,\mu) -  \frac{ 2 \pi i N}{2N} (\mu,w_2) - 2 \pi i\frac{N\chi}{24}  \right)\,\widehat h_{N,\mu}(\tau) \\
&= \exp{\left(
  \frac{ 2\pi i}{N}   q_{w_2}(\mu)-2\pi i\frac{N}{8}(w_2,w_2)
    - 2 \pi i\frac{N\chi}{24}   \right)}\,\widehat h_{N,\mu}(\tau) \\
&= \exp{\left(
  \frac{ 2\pi i}{N}   q_{w_2}(\mu)
    - 2 \pi i\frac{N}{24} (\chi + 3\sigma)   \right)}\,\widehat h_{N,\mu}(\tau) ,
\end{aligned}
\end{equation}}
where $K_S = Nw_2 \in H^2(S,\bZ)$ is the canonical class of $S$ and  $q_{w_2}$ is the quadratic refinement \eqref{eq:defq}.

The overall factor should be proportional to the left-moving central charge. By definition in \eqref{eq:genEG}, by the shift $\tau \to \tau +1$, the generalized elliptic genera will pick up a phase $e^{\frac{2\pi i}{24}c_L}$. From the anomaly polynomial \eqref{eqn:c20}, the left-moving central charge is 
\begin{equation}
    c_L =(2\chi+3\sigma)(N^3-N)+ N\chi
\end{equation}
We find that the first term, representing the  interaction between coincident M5 branes, disappears because the setup we consider is rigid when embedded into the Calabi-Yau background. The second term $N\chi$ from the $N$ free M5 branes matches exactly with the overall factor in front of the VW partition function \cite{Vafa:1994tf}.

By these observations, we try to relate the rank $N$ VW partition functions on $S$ to the generalized elliptic genera of $T_N[S]$.
In the rest of this work, we will use the known results on the VW partition functions to give the partition function or generalized elliptic genera of $T_N[S]$ on Hirzebruch and Del Pezzo surfaces. 

\section{M5 branes on Hirzebruch surface}
\label{sec:Hirze}

Consider $N$ M5 branes on $T^2 \times \mathbb{F}_l$. The dimensional reduction along the Hirzebruch surface $\bF_l$ gives a 2d $\cN=(0,4)$ theory denoted by $T_N[\mathbb{F}_l]$. Using our proposal in \eqref{eq:2d/4d}, we will use the VW partition function in \cite{Manschot:2011dj} to study the partition functions of different absolute theories of $T_N[\mathbb{F}_l]$.

Denoting the fiber and base divisor classes of the Hirzebruch surface $\bF_l$ by $f$ and $b$, one finds the intersection form
\begin{equation}
Q=\begin{pmatrix} 
f\cdot f & f\cdot b\\ b\cdot f & b\cdot b
\end{pmatrix}=\begin{pmatrix} 0 & 1\\1 & -l\end{pmatrix}.
\end{equation}
The canonical class of $\bF_l$ is $K_{\bF_l}= -(l+2) f -2 b$. The second Stielf-Witnney class is 
\begin{equation}
    w_2 = (l \textrm{ mod } 2,0) .
\end{equation}
Thus, $\bF_l$ is spin for even $l$ and non-spin when $l$ is odd. The Euler characteristic is $\chi =4$ and the signature $\sigma = 0$.

\paragraph{Coupling from geometry}

There is a coupling in the theory $T_N[\bF_l]$. We will determine it from the invariant volume of $\bF_l$ and compare the result with the expression of the VW partition function later.
Let us denote the K\"ahler class (Poincar\'{e} dual to the K\"ahler form) as
\begin{equation}
J=x f+y b\equiv 
\left(\begin{matrix} x \\ y \end{matrix}\right).
\end{equation}
The volume of the 4-manifold $\bF_l$ is thus given by
\begin{equation}
V_{\bF_l}=\frac{1}{2}J^T Q J\,.
\end{equation}
$V_{\bF_l}$ is invariant under the base change $J\rightarrow PJ$, where $P\in GL(2,\bZ)$ satisfies
\begin{equation}
P^T Q P=Q\,.
\end{equation}
Hence we conclude that the action of $P$ on the geometry of $\bF_l$ is exactly given by $J'=PJ$. We introduce the parameter
\begin{equation}\label{eq:defR}
    R=\frac{x}{y},
\end{equation}
which transforms non-trivially under the action of $\MCG(\bF_l)$. To see its geometric meaning, we compute the volume of 2-cycles
\begin{equation}
\begin{aligned}
V_f&=J\cdot f=y\cr
V_b&=J\cdot b=x- l y\cr
V_{b+lf}&=V_{lf}+V_b=x\,.
\end{aligned}
\end{equation}
Hence $R$ is the ratio of the volume of $f$ over the volume of $b+lf$, which are both 2-spheres,
\begin{equation}
    R=\frac{V_{b+lf}}{V_f}\,.
\end{equation}
In terms of $m$ and $n$, one has 
\begin{equation}\label{eq:defxy}
    y = m,\qquad x = ml+n.
\end{equation}
One can show that the choice $n=0$, giving a ratio
\begin{equation}
    R = l,
\end{equation}
is invariant under the action of the mapping class group.

\subsection{SymTFT}

After compactification, the 7d TQFT in equation \eqref{eqn:symTFT-7d} becomes 
\begin{eqnarray} \label{Eqn:actionF1}
    S_{3d} &=& \frac{N}{2\pi} \int \hat{a} \wedge da  - \frac{Nl}{4\pi} \int a \wedge d a,
\end{eqnarray}
where $a = \int_{b} c$ and $\hat{a} = \int_{f} c$. 
The corresponding $K$-matrix is given by 
\begin{equation}
    K = \begin{pmatrix}
        0 & N \\ N & -l N
    \end{pmatrix}, \quad
    K^{-1} = \begin{pmatrix}
        l/N & 1/N \\ 1/N & 0
    \end{pmatrix}.
\end{equation}
We see that the defect group is $\mathscr{D}_{\mathbb{F}_l} = \mathbb{Z}_N \times \mathbb{Z}_N$.
The topological spin is 
\begin{equation} 
\begin{aligned}
    \theta(\tilde \mu) = \exp{\left[ \frac{2\pi i }{N} q_{w_2}(\tilde \mu) \right]}
     = \exp{\left[\frac{2\pi i(1-N)}{2N} (2\mu_1 \mu_2 -l \mu_2^2) \right]},
\end{aligned}
\end{equation}
%
and the braiding matrix is 
\begin{equation} \label{eq:BFl}
\begin{aligned}
   B( \tilde \mu, \tilde \nu) =
   \exp\left[\frac{2\pi i}{N} (\mu_1 \nu_2+\mu_2 \nu_1-l \mu_2 \nu_2) \right].
\end{aligned}
\end{equation}
The modular transformations are determined to be
\begin{equation}\label{eq:STSymTFTFl}
\begin{aligned} 
    S(\tilde \mu, \tilde \nu) &= \frac{1}{N} \exp\left(\frac{2\pi i}{N} (\mu_2  \nu_1 +  \mu_1 \nu_2 + l \mu_2 \nu_2)\right),  \\
    T(\tilde \mu, \tilde \nu) &= \delta_{\mu\nu} \exp \left( \frac{2\pi i(1-N)}{2N}\left(2 \mu_1 \mu_2-l \mu_2^2\right) \right).
\end{aligned}
\end{equation}
%


\subsection{Vafa-Witten partition function}

We will collect results on Vafa-Witten partition functions for $N=2\textrm{ and 3}$ from \cite{Manschot:2011dj}. 
Let $J_{mn}=m(b+l f)+nf \in H^2(\bF_l,\bR)$ with $m,n>0$ be the K\"ahler form and $\mu=\beta b - \alpha f$ labels the t`Hooft magnetic flux. The VW partition functions for 2 $M5$ branes  are given by 
\begin{align}\label{eq:VWHirzebruch}
    h_{N=2,\mu}(\tau,z;\bF_l,J)
    &=\left(\frac{i}{\theta_1(\tau,2z)\eta(\tau)}\right)^{2}f_{2,\mu}(\tau,z;\bF_l,J_{mn}),
\end{align}
where $q=e^{2\pi i\tau}$, $w=e^{2\pi i z}$, and 
\begin{equation}
    f_{2,\beta b -\alpha f}(\tau,z;\bF_l,J_{mn})  =
A_{l,(\alpha,\beta)}(\tau,z)+\vartheta_{l,(\alpha, \beta)}^{m,n}(\tau,z)\quad
\alpha,\beta \in\{0,1\}.
\end{equation}
Here $A_{l,(\alpha,\beta)}$ are the Appell functions given by 
\begin{equation}
\begin{aligned}
A_{l,(0,0)}(\tau,z) &=-\frac{1}{2}\sum_{n\in \mathbb{Z}}q^{ln^2}w^{2(l-2)n}
+\sum_{n\in\mathbb{Z}}\frac{q^{ln^2}w^{2(l-2)n}}{1-q^{2n}w^4}
+\frac{i\,\eta(\tau)^3}{\theta_1(\tau,4z)},\\
A_{l,(0,1)}(\tau,z)& =-\frac{1}{2}\sum_{n\in \mathbb{Z}}q^{\frac{l}{4} (2n+1)^2}w^{(l-2)(2n+1)}
+q^{\frac{l}{4}}w^{l-2}\sum_{n\in\mathbb{Z}}\frac{q^{l n(n+1)}w^{2(l-2)n}}{1-q^{2n+1}w^4},\\
A_{l,(1,1)}(\tau,z)&=q^{\frac{l+2}{4}}w^{l}\sum_{n\in
\mathbb{Z}}\frac{q^{l n(n+1)+n}w^{2(l-2)n}}{1-q^{2n+1}w^4}, \\ 
A_{l,(1,0)}(\tau,z)&=w^2\sum_{n\in\mathbb{Z}}\frac{q^{ln^2+n}w^{2(l-2)n}}{1-q^{2n}w^4}
+\frac{i\,\eta(\tau)^3}{\theta_1(\tau,4z)},
\end{aligned}
\end{equation}
and the theta functions $\vartheta_{l,(\alpha, \beta)}^{m,n}(z,\tau)$ capture the K\"aher dependence
\begin{equation}\begin{aligned}
\vartheta_{l,(\alpha, \beta)}^{m,n}(\tau,z)&=\sum_{a,b \in \mathbb{Z}}\frac{1}{2}
\left(\sgn(-(2a-\alpha))-\sgn((2b-\beta)n-(2a-\alpha)m) \right)
\\
&\times w^{(l-2)(2b-\beta)+2(2a-\alpha)}\,q^{\frac{l}{4}(2b-\beta)^2+\frac{1}{2}(2b-\beta)(2a-\alpha)}.
\end{aligned}\end{equation}
The Appell functions $A_{l,(\alpha,\beta)}$ are examples of mock modular forms \cite{Zwegers:2008}.
Although they are holomorphic functions of $\tau$, they do not transform as a modular form.
One needs to complete them by small but necessary non-holomorphic terms called \textit{shadows} to render the transformation modular. 
The completion of $A_{l,(\alpha,\beta)}$ is
\begin{equation}\label{eq:Ahat}
\small
\begin{aligned}
\widehat
A_{l,(\alpha,\beta)}(\tau,z)=&A_{l,(\alpha,\beta)}(\tau,z)+\frac{1}{2}\sum_{k=0}^{l-1}\left(
  \sum_{ 
    n_1=2k+\beta l+\alpha \atop \mod 
    2l}w^{\frac{l-2}{l}n_1}q^{\frac{n_1^2}{4l}} \right)\\
&\times \sum_{n_2=-2k-\alpha \atop \mod 2l}\left[
  \sgn(n_2)-E\left(\left(n_2+2(l+2)\frac{\im z}{y}\right)\sqrt{\frac{y}{l}}\right)\right]
w^{-\frac{l+2}{l}n_2}q^{-\frac{n_2^2}{4l}},
\end{aligned}
\end{equation}
with $y=\im \tau$ and $E(x)=2\int_0^x e^{-\pi u^2}du$. Similarly, one can determine the completion of the theta function $\vartheta_{l,(\alpha, \beta)}^{m,n}$. The result is given by 
\begin{equation}\label{eq:thetahat}
\small
\begin{aligned}
\widehat{\vartheta}_{l,(\alpha, \beta)}^{m,n}(\tau,z)=&\vartheta_{l,(\alpha, \beta)}^{m,n}(\tau,z)
+\sum_{a,b \in \mathbb{Z}}\frac{1}{2}
\left[ E\left((-2a+\alpha+2(l+2)\frac{\im
    z}{y})\sqrt{\frac{y}{l}}\right) \right.   \\ 
 &-\left. E\left(((2b-\beta)n-(2a-\alpha)
  m+2(2n+(l+2)m)\frac{\im z}{y})
  \sqrt{\frac{y}{J^2_{m,n}}}\right)\right] \\ 
&\times
w^{(l-2)(2b-\beta)+2(2a-\alpha)}\,q^{\frac{l}{4}(2b-\beta)^2+\frac{1}{2}(2b-\beta)(2a-\alpha)},
\end{aligned}
\end{equation}
where $J^2_{m,n}=m(l m+2n)$.
After the modular completion, the VW partition functions are 
\begin{align} \label{eq:hhatFl}
    \widehat h_{2,\mu}(\tau,z;\bF_l,J)
    &=\left(\frac{i}{\theta_1(\tau,2z)\eta(\tau)}\right)^{2} \left(\widehat A_{l,(\alpha, \beta)}+\widehat \vartheta
_{l,(\alpha, \beta)}^{m,n}\right) \;.
\end{align}
They transform under $SL(2,\mathbb{Z})$ as 
\begin{equation}\label{eq:modtrA}
\begin{aligned}
\widehat h_{l,(\mu_1,\mu_2)} \left(\frac{-1}{\tau},\frac{z}{\tau}\right)& =\frac{1}{2\tau}e^{2\pi i(- \frac{12 \, z^2}{\tau})} 
\sum_{\nu_1,\nu_2 \in \{0,1\}}  
(-1)^{-l \mu_2 \nu_2 +\mu_1 \nu_2+\mu_2 \nu_1} \widehat h_{l,(\nu_1,\nu_2)}(\tau,z ),\\
\widehat h_{l,(\mu_1,\mu_2)}
\left(\tau+1,z\right)\,&= e^{2\pi i( \frac{l\mu_2^2-2\mu_1\mu_2}{4}-\frac{8}{24})}\,
\widehat h_{l,(\mu_1,\mu_2)} (\tau,z)\;.
\end{aligned}\end{equation}
Hence they are vector-valued Jacobi forms of weight $w=-1$ and index $m=-12$.
Here, we have performed the replacements $\mu_1 \to -\alpha$ and $\mu_2 \to \beta$ compared with the result in \cite{Manschot:2011dj}. Up to some $\mu$ independent factors, the above transformations match with the S-matrix and T-matrix \eqref{eq:STSymTFTFl} derived from the SymTFT.

Next, let's check our proposal for VW partition functions of $N=3$ on $\bF_l$.
Partial results for the VW partition function are given by 
\begin{equation}
\small
\begin{aligned}
f_{3,\beta C-\alpha f}(\tau,z;\bF_l, J_{m,n})=&-\sum_{a,b\in \mathbb{Z}} \frac{1}{2} (\, \sgn((3b-2\beta)n-(3a-2\alpha )m)-\sgn(3b-2\beta)
\,) \\
&\times \left(w^{(l-2)(3b-2\beta)+2(3a-2\alpha)}-w^{
    -(l-2)(3b-2\beta)-2(3a-2\alpha)}
\right)\\
&\times\,
q^{\frac{l}{12}(3b-2\beta)^2+\frac{1}{6}(3b-2\beta)(3a-2\alpha)} \\
& \times\,f_{2,bC-af}(z,\tau;\bF_l,J_{|3b-2\beta|,|3a-2\alpha|}),
\end{aligned}
\end{equation}
with $\beta = 1, 2 \textrm{ mod } 3$ and $\alpha\in \bZ$. As in the rank 2 case, $f_{3,\beta C-\alpha f}(\tau,z)$ is a mock modular form with holomorphic anomaly under $S$-transformation. We will not study their modular completion in this work.
Instead, we will focus on the $T$-transformation which does not require a modular completion. The VW partition functions are 
\begin{equation}
    h_{3,\mu}(\tau,z;\bF_l,J)
    =\left(\frac{i}{\theta_1(\tau,2z)\eta(\tau)}\right)^{3}f_{3,\mu}(\tau,z;\bF_l,J). 
\end{equation}
From this explicit expression, we find that
\begin{equation}
    h_{3,\mu}(\tau+1,z;\bF_l, J)=\exp{\left[-\frac{2\pi i}{3}(2\mu_1\mu_2-l\mu_2^2) - \frac{2\pi i}{24} \times 12 \right]}h_{3,\mu}(\tau,z;\bF_l, J).
\end{equation}
This agrees with the T-matrix obtained from the SymTFT for $N=3$ in equation \eqref{eq:STSymTFTFl} up to an overall phase that accounts for the chiral central charge of the 2d theory $T_3[\bF_l]$.

\subsection{Example: $\bF_1$} \label{subsec:F1}

The SymTFT in this case is the twisted $\bZ_N$ gauge theory. The action is given in equation \eqref{Eqn:actionF1} with $l=1$. The absolute theories or global variants have been studied in \cite{Chen:2023qnv}. We will focus on the $N=2$ case and give the partition function of the absolute theories $T_2[\bF_1]$.  

The SymTFT of the $N=2$ case is the double-semion (DS) model \cite{Ji:2019eqo} with the following 4 anyons 
\begin{equation}
    1:\; (0,0)\qquad b:\; (1,0)\qquad s:\; (0,1)\qquad \bar{s}:\; (1,1)\;. 
\end{equation}
Their topological spins are $\theta(1)=\theta(b)=1$, $\theta(s)=i$ and $\theta(\bar{s})=-i$.
From the braiding matrix in equation \eqref{eq:BFl}, we find one topological boundary condition 
\begin{equation} \label{eq:LF1}
   L=\{(0,0),\quad (1,0)\}\rightarrow \bZ_2. 
\end{equation}
So, there is only one absolute theory $\bZ_2$ according to its symmetry.
This anomaly can be probed from the braiding between lines in the SymTFT.

The modular transformation of the  SymTFT is the same as the DS topological order given by 
\begin{equation}\label{eq:DSmodular}
S= \frac{1}{2}\begin{pmatrix*}[r]
 1 & 1 & 1 & 1 \\
 1 & 1 & -1 & -1 \\
 1 & -1 & -1 & 1 \\
 1 & -1 & 1 & -1 \\
\end{pmatrix*},
\quad
T=\begin{pmatrix*}
1\\&1\\
&&i\\
&\phantom{-i}&\phantom{-i}&-i
\end{pmatrix*}.
\end{equation}
We find only one boundary 
\begin{equation}
    B=\left(
\begin{array}{c}
 1 \\
 1 \\
 0 \\
 0 \\
\end{array}
\right),
\end{equation}
which corresponds to the maximal sub-lattice we found. The modular invariant partition function is 
\begin{equation}
    Z[0,0] = Z_1 + Z_b,\quad Z[1,0] = Z_s + Z_{\bar s}.
\end{equation}
By performing $S$ and $T$ transformations, partition functions with TDL inserted along the space direction are 
\begin{equation}
    Z[0,1] = Z_1 - Z_b,\quad Z[1,1] = i Z_s -i Z_{\bar s}.
\end{equation}
One can identify the VW partition function with the SymTFT partition vector as 
\begin{equation}
\begin{aligned}
     & Z_1=\widehat h_{2,(0,0)} (\tau,z;\bF_1,J), \quad  Z_b=h_{2,(0,1)} (\tau,z;\bF_1,J), \\ 
     &Z_s=\widehat h_{2,(1,0)}(\tau,z;\bF_1,J) , \quad   Z_{\bar s}=h_{2,(1,1)}(\tau,z;\bF_1,J).
\end{aligned}
\end{equation}
The partition function of the absolute theory is 
\begin{equation} \label{eq:PFF1}
    Z_T=\widehat h_{2,(0,0)} (\tau,z;\bF_1,J) + \widehat h_{2,(0,1)} (\tau,z;\bF_1,J).
\end{equation}
After making it modular invariant, it is expected to be the partition function of the 2d $\cN=(0,4)$ theory $T_2[\bF_1]$. 

\subsection{Example: $\bF_2$} \label{subsec:F2}

The SymTFT for $\bF_2$ case has the action 
\begin{equation}
    S_{3d} = \frac{N}{2\pi} \int \hat{a} \wedge da  - a \wedge d a.
\end{equation}
In contrast to the twisted $\bZ_N$ gauge theory studied before, it is a $\bZ_N$ gauge theory by a field redefinition $\hat{a} - a \to \hat{a}$. 
The defect group is $(\bZ_N)^2$. The possible absolute theories or global variants have been studied in \cite{Chen:2023qnv}.
We will again focus on the case of $N=2$ and give the partition function of the absolute theories $T_2[\bF_2]$
\footnote{Since some of the VW partition functions on $\bF_0$ are not well defined, we will study them on $\bF_2$ instead. The absolute theories labeled by the maximal isotropic sub-lattice are the same as those for $\bF_0$ up to a basis transformation.}.

The SymTFT for $N=2$ is the toric code. There are 4 anyons 
\begin{equation}
    1:\; (0,0)\qquad e:\; (0,1)\qquad m:\; (1,0)\qquad f:\; (1,1)\;. 
\end{equation}
Their topological spins are $\theta(1)=\theta(e)=\theta(m)=1$ and $\theta(f)=-1$. From Eq. \eqref{eq:BFl}, we find the following three maximal isotropic sub-lattices,
\begin{equation} \label{Eqn:LN2}
\begin{array}{cccc}
L_1= \{(0,0), & (1,0)\} &\to &\bZ_2  \\
L_2 = \{ (0,0), & (1,1)\}&\to &\widehat{\bZ}_2 \\
L_3=\{ (0,0), & (0,1)\} &\to &\;\;\bZ_2^f\;. 
\end{array}
\end{equation}
Each one of them corresponds to an absolute theory. Since they all have $\bZ_2$ symmetry, we will label them as $\bZ_2$, $\widehat{\bZ}_2$, and $\bZ_2^f$. Besides that, one can stack them with the Arf invariant. Let's denote the theory with or without this SPT phase by the subscript $0/1$.

There are in total 6 global variants. They form a groupoid determined by the automorphism group $\Aut_{\bZ_2}(Q)=S_3$. The two generators of $S_3$ are 
\begin{equation} \label{eq:topActF2}
\sigma=\begin{pmatrix}
1 & 0 \\
 1 & 1 \\
\end{pmatrix},
\quad 
\tau=\begin{pmatrix}
1 & 1 \\
 0 & 1 \\
\end{pmatrix},
\end{equation}
which can be identified as the following two topological operations. 
\begin{itemize}
    \item  $\sigma$ represents gauging of $\bZ_2$. In terms of the partition function, it is given by 
\begin{eqnarray}\label{eq:2dgauging}
 \widehat Z[A] =
 \frac{1}{2} 
 \sum_{a\in H^1(T^2, \bZ_2)} Z[a]\,(-1)^{\int a \cup A} ,
\end{eqnarray}
where $a$ is the background field of $\bZ_2$ in the original theory and $A$ is the background field of the quantum symmetry $\widehat\bZ_2$ after gauging.
\item 
$\tau$ represents stacking with the SPT phase. 
The partition function after stacking with the SPT phase is simply \cite{Ji:2019eqo}
\begin{equation}
    Z[A]= (-1)^{\Arf[A+\rho]+\Arf[\rho]} Z[A],
\end{equation}
where $\rho$ is a choice of spin structure and $A$ is the $\bZ_2$ background gauge field. 
\end{itemize}

\begin{figure}[ht]
\centering
\begin{tikzpicture}[scale=1.5]
\draw node at (0,0) {$(\bZ_2)_0$};
\draw node at (3,0) {$(\widehat \bZ_2)_0$};
\draw node at (6,0) {$(\bZ^f_2)_0$};
\draw node at (0,2) {$(\bZ_2)_1$};
\draw node at (3,2) {$(\widehat \bZ_2)_1$};
\draw node at (6,2) {$(\bZ_2^f)_1$};
\draw [<->,orange] (.5,0) -- (2.5,0);
\draw [<->,orange] (.5,2) -- (2.5,2);
\draw [<->,blue] (0.05,0.2) -- (0.05,1.8);
\draw [<->,blue] (3,0.2) -- (3,1.8);
\draw [<->,blue] (5.95,0.2) -- (5.95,1.8);
\draw [<->,orange] (6.05,0.2) -- (6.05,1.8);
\draw [<->,blue] (0.2,1.8) -- (5.8,0.2);
\draw [<->,blue] (0,-.2) arc (-120:-60:3);
\draw [<->,blue] (3,2.2) arc (120:60:3);
\draw node at (1.5,.2) {${\color{orange} s}$};
\draw node at (1.5,2.2) {${\color{orange} s}$};
\draw node at (1.5,-.45) {${\color{blue} \sigma}$};
\draw node at (4.5,2.75) {${\color{blue} \sigma}$};
\draw node at (0.2,1) {${\color{blue} \tau}$};
\draw node at (5.8,1) {${\color{blue} \tau}$};
\draw node at (6.2,1) {${\color{orange} s}$};
\draw node at (2.8, 0.6) {$\color{blue} \tau$};
\draw node at (4, 0.9) {$\color{blue} \sigma$};
\node[below] at (6,-0.5) {$\begin{pmatrix}
0 & 1 \\ 1 & 0
\end{pmatrix}$};
\node[above] at (6,2.5) {$\begin{pmatrix} 0 & 1 \\ 1 & 1 \end{pmatrix}$};
  \node[below] at (3,-0.5) {$\begin{pmatrix} 1 & 0 \\ 0 & 1 \end{pmatrix}$};
 \node[above] at (3,2.5) {$\begin{pmatrix} 1 & 1 \\ 0 & 1 \end{pmatrix}$};
 \node[below]  at (0,-0.5) {$\begin{pmatrix} 1 & 0 \\ 1 & 1 \end{pmatrix}$};
 \node[above] at (0,2.5) {$\begin{pmatrix} 1 & 1 \\ 1 & 0 \end{pmatrix}$};
\end{tikzpicture}
\caption{Web of transformations for $T_2[\bF_2]$. The transformations
in orange are the duality transformations.
The transformations in blue are topological manipulations.}
\label{Fig:N2f2}
\end{figure}
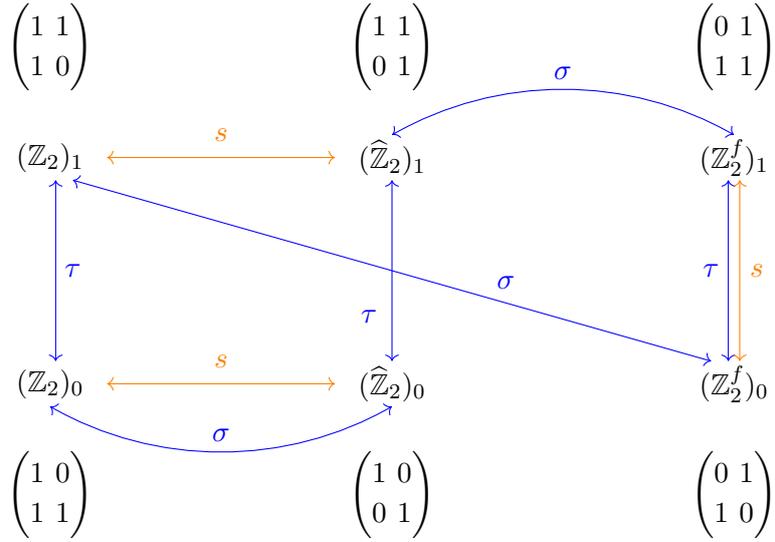

There is a correspondence between these global variants and the automorphism group \cite{Gukov:2020btk, Gaiotto:2020iye}. It is convenient to assign each global variant to a two-dimensional representation $\{M_i\}_{i=1,2,\ldots, 6}$ of $\Aut_{\bZ_2}(Q)$. The transformation between these global variants under the 
topological operations $\sigma$ and $\tau$ are simply determined by acting on the $\{M_i\}$ with the matrix representations \eqref{eq:topActF2} from the right. We give an example of such an assignment in Figure \ref{Fig:N2f2} where $s$ is a non-trivial $Q$ preserving automorphism given by 
\begin{equation} \label{eq:sF2}
    s = \begin{pmatrix*}[r]
        1 & 0 \\
        1 & -1
    \end{pmatrix*}.
\end{equation}
It generates a $\bZ_2$ actions on the $\{M_i\}$ by multiplying \eqref{eq:sF2} from the left and can be understood as a duality among different global variants of $T_2[\bF_2]$. 

%

%
The S- and T-matrices of the toric code can be determined from \eqref{eq:STSymTFTFl} as 
\begin{equation}\label{eq:modularF2}
S=\frac{1}{2}\begin{pmatrix*}[r]
 1 & 1 & 1 & 1 \\
 1 & 1 & -1 & -1 \\
 1 & -1 & 1 & -1 \\
 1 & -1 & -1 & 1 \\   
\end{pmatrix*}, \quad
T = \begin{pmatrix}
 1 & \phantom{-1} & \phantom{-1} & \phantom{-1} \\
  & 1  \\
  &  & 1 \\
  &  &  & -1 \\   
\end{pmatrix}.
\end{equation}
We find three boundaries
\begin{equation}
    B_1=\begin{pmatrix}
        1\\1\\0\\0
    \end{pmatrix},\quad
    B_2=\begin{pmatrix}
        1\\0\\1\\0
    \end{pmatrix},\quad
    B_3=\begin{pmatrix}
        1\\0\\0\\1
    \end{pmatrix},
\end{equation}
corresponding to three maximal sub-lattices in \eqref{Eqn:LN2}.

Consider the first two polarizations labeled by $L_1$ and $L_2$. Since anyons inside both of these two polarizations have spin one, the theories on the boundary are bosonic. 
The partition functions of $L_1$ and $L_2$ are given by 
\begin{equation} \label{eq:l1F2}
     Z[0,0] = Z_1 + Z_e 
\end{equation}
and 
\begin{equation}\label{eq:l2F2}
    \widehat Z[0,0] = Z_1 + Z_m.
\end{equation}
%
These two partition functions are related by the electric-magnetic duality $e \leftrightarrow m$ in the SymTFT or by gauging the $\bZ_2$ symmetry on the boundary.  This is consistent with the transformation between global variants shown in Figure \ref{Fig:N2f2}.

Next, let's consider the absolute theory specified by $L_3$. Since this polarization contains the fermionic line, i.e. an anyon with topological spin $\theta(f)=-1$, the absolute theory is a fermionic theory. The partition functions with respect to different spin structures are given by \cite{Ji:2019eqo}
\begin{equation}
\begin{aligned}
& Z^f[AA] = Z_1 + Z_f ,\qquad Z^f[PA] = Z_e + Z_m,\\
&Z^f[AP] = Z_1 - Z_f ,\qquad Z^f[PP] = Z_e - Z_m,
\end{aligned}
\end{equation}
where $P$ and $A$ are periodic and anti-periodic boundary conditions.

By comparing Eq. \eqref{eq:modularF2} and Eq. \eqref{eq:STSymTFTFl}, the conformal blocks of the SymTFT transform in the same way as the VW partition functions of $\bF_2$. 
We will associate them with the corresponding VW partition functions as follows 
\begin{equation} \label{eq:hzF2}
\begin{aligned}
    &Z_1 = \widehat h_{2,(0,0)}(\tau,z;\bF_2,J) , \quad 
    Z_e = \widehat h_{2,(0,1)} (\tau,z;\bF_2,J), \\\
    &Z_m = \widehat h_{2,(1,1)}(\tau,z;\bF_2,J) , \quad Z_f = \widehat h_{2,(1,0)}(\tau,z;\bF_2,J)  \;.
\end{aligned}
\end{equation}
Up to a non-trivial modular weight, we obtain the partition functions of the three absolute theories. In particular, we identify a particular combination of VW partition functions that transforms as a fermionic partition function under the modular group.

\paragraph{Duality defect in $\bF_2$ theory}
We have seen that absolute theories transform into each other under the $s$ operation defined in \eqref{eq:sF2}.
In particular, it switches between the theories labeled by $(\bZ_2)_0$ and $(\widehat \bZ_2)_0$.
Following the arrows in Figure \ref{Fig:N2f2}, one can see that the operation $\sigma s$ is an identity operation on the theory $(\bZ_2)_0$.
The only difference is that the coupling constant $R$ after the s operation becomes 
\begin{equation}
    s(R)=\frac{R}{R-1}. 
\end{equation}
If one takes the coupling to be $R=2$, the composition of $s$ and $\sigma$ transform from $(\bZ_2)_0$ to itself exactly and define a duality defect \cite{Choi:2021kmx,Chen:2023qnv}. 

The existence of the duality defect implies that the partition function of the theory is invariant under the gauging of the $\bZ_2$ 0-form symmetry \cite{Chang:2018iay}. 
If the 2d partition function can be correctly identified with the VW one, we expect that the corresponding VW partition function is also invariant under $\sigma$. 
Let's focus on the absolute theory $(\bZ_2)_0$. The modular invariant partition function in terms of the VW partition function is given in \eqref{eq:l1F2}. Again from the Figure \ref{Fig:N2f2}, the gauging operation transforms the theory from $(\bZ_2)_0$ to another absolute theory $(\widehat \bZ_2)_0$. The partition function of $(\widehat \bZ_2)_0$ is given in \eqref{eq:l2F2}. It is expected that the partition function before and after gauging should be the same for the value of $J$ where $R=2$, i.e.    
\begin{equation}
    \widehat h_{2,(1,0)}(\tau,z;\bF_2,J) =
   \widehat h_{2,(1,1)} (\tau,z;\bF_2,J).
\end{equation}
We check this statement using the explicit expressions. The details can be found in Appendix \ref{App2}.


\section{M5 branes on Del Pezzo surfaces}
\label{sec:delPezzo}

In this section, we will consider the compactification of 6d $\cN=(2,0)$ SCFTs of type $A_{N-1}$ on the Del Pezzo surface $dP_l$. Let's denote the 2d $\cN=(0,4)$ SCFTs as $T_N[dP_l]$. We will study the global variants and symmetries of $T_N[dP_l]$ using the SymTFT. Also, with the known results for $\bP^2$ \cite{Bringmann:2010sd,Manschot:2010nc,Manschot:2014cca,Manschot:2021qqe,Chattopadhyaya:2023aua}, we calculate the rank two VW partition function of $dP_l$ with $l>0$ using the blow-up formula \cite{Manschot:2011ym,Haghighat:2011xx,Haghighat:2012bm} and based on our proposal in \eqref{eq:2d/4d}, relate the result to the partition vector (functions) of $T_N[dP_l]$. 

The Picard group generators of $dP_l$ are denoted as $h,e_1,e_2, \ldots, e_l$ satisfying the intersection relations
\begin{equation}
    h^2=1\ ,\ h\cdot e_i=0\ ,\ e_i\cdot e_j=-\delta_{i,j}\quad (i,j=1,2,\ldots,l)\,.
\end{equation}
The intersection form of $dP_l$ is a rank $(l+1)$ matrix, of the form
\begin{equation}
    Q_{ij}=\text{diag}(1,-1,\dots,-1)\,.
\end{equation}
The canonical class of $dP_l$ is $K_{S} = -3 h + \sum_{i=1}^l e_i$. The second Stiefel-Whitney class is 
\begin{equation}
    w_2 =  (1,1,\dots,1),\quad \mod \; 2.
\end{equation}
Thus, all $dP_l$ with $l>0$ are non-spin manifolds.
The Euler characteristic and the signature are
\begin{equation}
    \chi(dP_l)=l+3,\quad \sigma(dP_l)=1-l.
\end{equation}

\subsection{SymTFT}

The SymTFT of the 2d theory $T_N[dP_l]$ is given by 
\begin{eqnarray} \label{Eqn:actiondp}
    S_{3d} &=& \frac{N}{4\pi} \int_{W_3} a_0 \wedge da_0  -  \frac{N}{4\pi}  \sum_{i=1}^{l} \int_{W_3} a_i \wedge da_i,
\end{eqnarray}
where the $K$-matrix is $K^{ij} = NQ^{ij}$. This is an abelian Chern-Simons theory which is bosonic for even $N$, and otherwise spin. The defect group is $\mathscr{D} = H^2(dP_l,\bZ_N)=(\mathbb{Z}_N)^{l+1}$.
%
%
%
The topological spin of an anyon $\tilde \mu$ is given by 
\begin{equation}
\begin{aligned}
    \theta(\tilde \mu) &= \exp{\left[ \frac{2\pi i }{N} q_{w_2}(\tilde \mu) \right]}= \exp{\left[ \frac{2\pi i }{2N} (\mu-\frac{N}{2}w_2, \mu-\frac{N}{2}w_2) \right]}\\
     & = \exp{\left[\frac{2\pi i(1-N)}{2N} ((\mu^0)^2- \sum_{i=1}^{l}(\mu^i)^2) + \frac{2\pi iN(1-l)}{8} \right]},
\end{aligned}
\end{equation}
where we have used equation \eqref{eq:m4fact1} here. 
The braiding between two different anyons $\tilde \mu$ and $\tilde \nu$ is 
\begin{equation} \label{eq:Bdelpezzo}
    B ( \tilde \mu,\tilde \nu) = \exp\left(\frac{2\pi i}{N} (\mu^0 \nu^0-\sum_{i=1}^{l}\mu^i \nu^i )\right).
\end{equation}
So, the S- and T-matrices are determined to be
\begin{equation}\label{eq:modularSymTFTdP}
\begin{aligned}
    S(\tilde \mu,\tilde \nu) & = \frac{1}{N^{(l+1)/2}} \exp\left(\frac{2\pi i}{N} (\mu^0 \nu^0-\sum_{i=1}^{l}\mu^i \nu^i )\right), \\
     T(\tilde \mu,\tilde \nu) & = \exp{\left[\frac{2\pi i(1-N)}{2N} ((\mu^0)^2- \sum_{i=1}^{l}(\mu^i)^2) + \frac{2\pi iN(1-l)}{8} \right]} \delta_{\mu\nu}.
\end{aligned}
\end{equation}
%
As we will see, except for the overall factor in the T-matrix, the above matrices match with the modular transformation properties of the VW partition functions on $dP_l$.

From the braiding matrix \eqref{eq:Bdelpezzo}, we can determine the possible topological boundary conditions from the equation \eqref{Eqn:TopBndCon}. For odd $l$, the discriminant group $\sD=(\mathbb{Z}_N)^{l+1}$ is a Drinfeld double. The topological boundary conditions exist and the corresponding absolute theory has 0-form symmetry $\bZ^{(l+1)/2}$. However, when $l$ is even, the SymTFT does not have topological boundary conditions and $T_N[S]$ is always relative.

\subsection{Example: $\bP^2$}

The SymTFT of $T_N[\bP^2]$ is the $U(1)_N$ Chern-Simons theory. The discriminate group $\sD=\bZ^N$ which implies that there are topological boundary condition and $T_N[\bP^2]$ is generically relative. According to our proposal in \eqref{eq:2d/4d}, we will relate the rank two VW partition function on $\bP^2$ with the partition vector of $T_2[\bP^2]$.

The rank two refined VW partition function on $\bP^2$ is 
\begin{equation} \label{eq:RVWP2}
    h_{2,\mu}(\tau,z)= g_{2,\mu}(\tau,z) \left( \frac{i}{\theta_1(\tau,2z)}\right)^2,
\end{equation}
with 
\begin{equation}
\small
g_{2,0}(\tau,z)=\frac{1}{2} +\frac{q^{-\frac{3}{4}}w^5}{\theta_2(2\tau,2z)}\sum_{n\in\bZ}\frac{q^{n^2+n}w^{-2n}}{1-w^4q^{2n-1}}, \quad
g_{2,1}(\tau,z) =\frac{q^{-\frac{1}{4}}w^3}{\theta_3(2\tau, 2z)}\sum_{n\in\bZ}\frac{q^{n^2}w^{-2n}}{1-w^4q^{2n-1}}.
\end{equation}
The VW partition function on $\bP^2$ for $N=2$ is obtained by taking the unrefined limit 
\begin{equation}
    h_{2,\mu}(\tau) :=\lim_{z\to 0}4\pi i z\, h_{2,\mu}(\tau,z)= \frac{f_{2,\mu}(\tau)}{\eta(\tau)^6},
\end{equation}
in which the holomorphic function $f_{2,\mu}(\tau)$ can be shown to be equal to
\begin{equation}
    f_{2,\mu}(\tau) = 3 \sum_{n\geq 0} H(4n-\mu) q^{n-\frac{\mu}{4}}\;.
\end{equation}
Here, $H(4n-\mu)$ is the generating function of Hurwitz class numbers\footnote{
The Hurwitz class number $H(n)$ is a modification of the class number of positive definite binary quadratic forms.
$H(1,2\textrm{ mod }4)=0$ and the first few terms of $H(n)$ are $\sum_{n=0}^{20}H(n)q^n=-\frac{1}{12}+\frac{1}{3}q^3+\frac{1}{2}q^4+q^7+q^8+q^{11}+\frac{4}{3}q^{12}+2q^{15}+\frac{3}{2}q^{16}+q^{19}+2q^{20}.$
}.
The functions $f_{2,\mu}(\tau)$ are mock modular forms.
After the completion, one has that 
\begin{equation}
    \widehat f_{2,\mu}(\tau)=f_{2,\mu}(\tau)-\frac{3i}{4\sqrt{2}\pi} \int_{-\bar \tau}^{i\infty}\frac{\Theta_{\mu/2}(u)}{(-i(\tau+u))^{\frac{3}{2}}}du,
\end{equation}
where $\Theta_\alpha$ is the theta series 
\begin{equation}
    \Theta_\alpha(\tau)=\sum_{k\in \bZ+\alpha} q^{k^2}, \quad q = e^{2\pi i \tau}.
\end{equation}
The modular property for $\widehat h_{2,\mu}$ is
\begin{equation}\label{eq:Sh20}
    \begin{pmatrix}
        \widehat{h}_{2,0}\\ \widehat{h}_{2,1}
    \end{pmatrix}\left(\frac{-1}{\tau}\right)
    =-\left(\frac{\tau}{i}\right)^{-\frac{3}{2}}\frac{1}{\sqrt{2}}
    \begin{pmatrix}
        1&1\\1&-1
    \end{pmatrix}\begin{pmatrix}
        \widehat{h}_{2,0}\\ \widehat{h}_{2,1}
    \end{pmatrix}(\tau).
\end{equation}
This is the desired transformation property by setting $l=0$ in Eq. \eqref{eq:modularSymTFTdP}.

Generally, for $N>2$, it is believed that the modular transformation of $h_{N,\mu}(\tau)$ takes the form
$ h_{N,\mu}(\tau)=f_{N,\mu}(\tau)/\eta(\tau)^{3N} $
and transforms as \cite{Manschot:2017}
\begin{equation}
\begin{aligned}
\widehat{h}_{N,\mu}\!\left(-\frac{1}{\tau} \right)&=\frac{(-1)^{N-1}}{\sqrt{N}}
\left(\frac{\tau}{i}\right)^{-\frac{3}{2}} \sum_{\nu=0}^{N} e^{-2\pi i
  \frac{\mu\nu}{N}} \widehat{h}_{N,\nu}(\tau),\\
\widehat{h}_{N,\mu}(\tau+1)&=e^{2\pi i(-\frac{N}{4}+\frac{1}{2N}(\mu+N/2)^2) } \widehat{h}_{N,\mu}(\tau)
=e^{2\pi i(\frac{\mu^2}{2N}+\frac{\mu}{2}-\frac{N}{8})} \widehat{h}_{N,\mu}(\tau).
\end{aligned}
\end{equation}
Up to the overall factor, it matches the results of the SymTFT in \eqref{eq:modularSymTFTdP}. Note that the $\mu$ independent factor from the T-transformation gives the partial left-moving central charge $N\chi = 3N$. 




\subsection{Example: $dP_1$}

Let's first consider $S=dP_1$ with intersection form 
\begin{equation}
    Q=\bp h \cdot h & h \cdot e \\ e \cdot h & e \cdot e \ep  = \bp 1 & 0\\0 & -1\ep .
\end{equation}
It is isomorphic to $\bF_1$ by the basis transformation 
\begin{equation} \label{eq:basisF1dp1}
    h = f + b, \qquad e = b\;.
\end{equation}
The SymTFT and corresponding absolute theories are the same as for the $\bF_1$ case discussed in Section \ref{subsec:F1}.
The VW partition functions of $dP_1$ are the same as those given in Eq. \eqref{eq:hhatFl} up to the basis transformation above.
Here, we will also derive it using the blow-up formula. This approach is easy to extend to $\bF_l$ for $l>1$.

The SymTFT of $T_N[dP_1]$ is 
\begin{eqnarray}
    S_{3d} &=& \frac{N}{4\pi} \int_{W_3} a_0 \wedge da_0  -  \frac{N}{4\pi}  \int_{W_3} a_1 \wedge da_1,
\end{eqnarray}
which is the same as the SymTFT of $T_N[\bF_1]$ by the field redefinition $a_0 \to a + \tilde a$ and $a_1 \to \tilde a$. The discriminant group is $\sD = \bZ_N \times \bZ_N$. For $N=2$, we find one topological boundary condition 
\begin{equation*}
   L=\{(0,0),\quad (1,1)\}\rightarrow \bZ_2\;.
\end{equation*}
It is the same as the one for $T_2[\bF_1]$ in \eqref{eq:LF1} by the basis transformation in equation \eqref{eq:basisF1dp1}.

Next, we will calculate the rank two VW partition functions of $dP_1$ and relate them to the partition functions of $T_2[dP_1]$. Given the VW partition function of $\bP^2$ in \eqref{eq:RVWP2}, one can obtain the partition function of $dP_1$ by the blow-up formula \cite{Manschot:2011ym,Haghighat:2011xx,Haghighat:2012bm}. Let $\tilde S$ be the surface from the blow-up $\phi: \tilde S \to S$ at a non-singular point. If $\gcd (N,(\mu,J)) = 1$, their VW partition functions are related by  
\begin{equation}
    h_{N,\phi^* \mu-k e}(z,\tau;\tilde S, \phi^* J)=B_{N,k}(z,\tau)\, h_{N,\mu}(z,\tau;S,J),
\end{equation}
where $e$ is the exceptional divisor and $B_{N,k}(z,\tau)$ is 
\begin{equation}
B_{N,k}(z,\tau)=\frac{1}{\eta(\tau)^N}\sum_{\sum_{i=1}^N a_i=0 \atop a_i\in \mathbb{Z}+\frac{k}{N}}
q^{\frac{1}{2}\sum_{i=1}^r a_i^2} w^{\sum_{i<j}a_i-a_j}.  
\end{equation}

When $N=2$, it is 
\begin{equation}
B_{2,k}(z,\tau)=\frac{1}{\eta^2(\tau)}\sum_{n\in \bZ+k/2}
q^{n^2} w^n,
\end{equation}
or in the notation of $\theta$ functions
\begin{equation}
    B_{2,0}(\tau,z)=\frac{\theta_3(2\tau,z)}{\eta(\tau)^2},\quad
    B_{2,1}(\tau,z)=\frac{\theta_2(2\tau,z)}{\eta(\tau)^2}.
\end{equation}
By using the theta function identity
\begin{equation}\begin{aligned}
\theta_3(\tau/2,z/2)&=\theta_3(2\tau,z)+\theta_2(2\tau,z),\\
\theta_4(\tau/2,z/2)&=\theta_3(2\tau,z)-\theta_2(2\tau,z),
\end{aligned}
\end{equation}
$B_{2,k}$ transforms under $\tau\to-1/\tau$ as
\begin{equation}
    \begin{pmatrix}
        B_{2,0}\\B_{2,1}
    \end{pmatrix}\to
    \sqrt{\frac{i}{\tau}}e^{\frac{\pi i z^2}{2\tau}}
    \sqrt{\frac{1}{2}}\begin{pmatrix}
        1&1\\1&-1
    \end{pmatrix}\begin{pmatrix}
        B_{2,0}\\B_{2,1}
    \end{pmatrix}.
\end{equation}

For $S=\bP^2$, the blowup formula gives us the VW partition functions of $dP_1$.
The modular completion of them is
\begin{equation}\begin{aligned}
    \widehat h_{2,-kb}(z,\tau;dP_1, \phi^* J)&=B_{r,k}(z,\tau)\, \widehat h_{2,0}(z,\tau;\bP^2,J), \quad k=0,1\\
     \widehat h_{2,f+(1-k)b}(z,\tau;dP_1, \phi^* J)&=B_{r,k}(z,\tau)\, \widehat h_{2,1}(z,\tau;\bP^2,J), \quad k=0,1.
\end{aligned}
\end{equation}
Note that $\phi^* h = f+b$ and $h$ is the hyperplane class of $\bP^2$. 
Taking the $S$ transformation of $\widehat h_{2,\mu}$ \eqref{eq:Sh20} into account, the overall $S$ matrix is (apart from the modular weight which is suppressed here)
\begin{equation}
    \begin{pmatrix*}[l]
        \widehat h_{2,0}\\\widehat h_{2,b}\\\widehat h_{2,f+b}\\\widehat h_{2,f}
    \end{pmatrix*}\to\frac{1}{2}\begin{pmatrix*}[r]
        1&1&1&1\\1&-1&1&-1\\1&1&-1&-1\\1&-1&-1&1
    \end{pmatrix*}\begin{pmatrix*}[l]
        \widehat h_{2,0}\\\widehat h_{2,b}\\\widehat h_{2,f+b}\\\widehat h_{2,f}
    \end{pmatrix*},
\end{equation}
and the $T$ matrix is
\begin{equation}
    T=\begin{pmatrix*}
        1\\&i\\
        &&-i&\phantom{-i}\\
        \phantom{-i}&\phantom{-i}&\phantom{-i}&1
    \end{pmatrix*}.
\end{equation}
Compared with the $S$ and $T$ matrix in Eq. \eqref{eq:DSmodular}, we can identify $\widehat h_{2,0},\widehat h_{2,b},\widehat h_{2,f+b},\widehat h_{2,f}$ with $1,s,\bar s,b$. Again, this also matches with the S- and T-matrices of the SymTFT in \eqref{eq:modularSymTFTdP}.

One can also check by explicitly expanding that the partition functions $h_2$ from the blowup formula agree with $\bF_1$'s partition function in Eq. \eqref{eq:VWHirzebruch}. Thus, the partition function of the absolute theory labeled by $L$ is the same as the one of $T_2[\bF_1]$ given in \eqref{eq:PFF1}.

\subsubsection*{Generalization to $dP_l$}

It is straightforward to generalize to the $dP_{l>1}$ case. The VW partition functions are
\begin{equation}
\begin{aligned}
    h_{2,(0,-k_1,-k_2,\ldots, -k_l)}(z,\tau;dP_l, \phi^* J)&=\textstyle\prod_{i=1}^l B_{2,k_i}(z,\tau)\, h_{2,0}(z,\tau;\bP^2,J),\\
     h_{2,(1,-k_1,-k_2,\ldots, -k_l)}(z,\tau;dP_l, \phi^* J)&=\textstyle\prod_{i=1}^l B_{2,k_i}(z,\tau)\, h_{2,1}(z,\tau;\bP^2,J), 
\end{aligned}
\end{equation}
where $k_i=0,1$ and the K\"ahler form $\phi^*h=h$.
It is easy to derive the $S$ and $T$ matrices to be
\begin{equation}
    S=\otimes^{l+1} \sqrt{\frac{1}{2}}\begin{pmatrix}
        1&1\\1&-1
    \end{pmatrix},\quad
    T=
    \begin{pmatrix}
        1\\&-i
    \end{pmatrix}\otimes^l \begin{pmatrix}
        1&\phantom{-i}\\&i
    \end{pmatrix}.
\end{equation}
%
%
This agrees with the SymTFT result in Eq. \eqref{eq:modularSymTFTdP}.

\subsection*{Acknowledgments}
We would like to thank Sergei Alexandrov, Azeem Hasan, Yihua Liu, and Michele Del Zotto for their valuable discussions. The work of BH and YS is supported by the NSFC grant 12250610187. The work of JC is supported by the Fundamental Research Funds for the Central Universities (No.20720230010) of China, and the National Natural Science Foundation of China (Grants No. 12247103). WC also would like to thank the organizers of the SymTFT Workshop at Peng Huanwu Center for Fundamental Theory (PCFT) for hospitality where part of this work was completed.

\appendix

\section{(Mock) Modular forms} \label{app1}


The Dedekind $\eta$ function is a modular form of weight $1/2$ and is defined by
\begin{equation}
\eta(\tau):=q^{\frac{1}{24}}\prod_{n=1}^\infty (1-q^n)
=\sum_{n=1}^{\infty}(-1)^{n}q^{(3n^2+n)/2}
\end{equation}
where $q=e^{2\pi i \tau}$ and transforms as
\begin{equation}
\eta(\tau+1) = e^{\frac{\pi i}{12}} \eta(\tau), \quad 
\eta(-\frac{1}{\tau}) = \sqrt{\frac{\tau}{i}}\,\eta(\tau).
\end{equation}
The Jacobi theta functions are defined by
\begin{equation}
\begin{aligned}
\theta_1(\tau,z):&=\sum_{n\in\bZ+\frac{1}{2}}(-1)^n q^{\frac{1}{2}n^2}w^n\\
\theta_2(\tau,z):&=\sum_{n\in\bZ+\frac{1}{2}} q^{\frac{1}{2}n^2}w^n\\
\theta_3(\tau,z):&=\;\;\sum_{n\in\bZ}\;\; q^{\frac{1}{2}n^2}w^n\\
\theta_4(\tau,z):&=\;\;\sum_{n\in\bZ}\;\;(-1)^n q^{\frac{1}{2}n^2}w^n\\
\end{aligned}
\end{equation}
where $q=e^{2\pi i \tau}$ and $w=e^{2\pi i z}$, with $\tau \in \mathbb{H}$ and $z\in \bC$. In the case $z=0$, $\theta_i(\tau,0)$ is denoted by $\theta_i(\tau)$.
Under the modular transformation, they transform as
\begin{equation}\begin{aligned}
    \theta_1(\tau+1,z) &= e^{\frac{\pi i}{4}}\theta_1(\tau,z), \quad 
    \theta_1(-1/\tau,z/\tau) = -i \alpha\theta_1(\tau,z)\\
    \theta_2(\tau+1,z) &= e^{\frac{\pi i}{4}}\theta_2(\tau,z), \quad 
    \theta_2(-1/\tau,z/\tau) = \alpha\theta_4(\tau,z)\\
    \theta_3(\tau+1,z) &= \theta_4(\tau,z), \quad 
    \theta_3(-1/\tau,z/\tau) = \alpha\theta_3(\tau,z)\\
    \theta_4(\tau+1,z) &= \theta_3(\tau,z), \quad 
    \theta_4(-1/\tau,z/\tau) = \alpha\theta_2(\tau,z)
\end{aligned}
\end{equation}
where $\alpha=\sqrt{\frac{\tau}{i}} e^{\frac{\pi i z^2}{\tau}}$.
The Jacobi theta functions satisfy a large number of identities.
For example,
\begin{equation}\begin{aligned}
\theta_3(\tau,z)&=\theta_3(4\tau,2z)+\theta_2(4\tau,2z),\\
\theta_4(\tau,z)&=\theta_3(4\tau,2z)-\theta_2(4\tau,2z).
\end{aligned}
\end{equation}


Mock modular forms are holomorphic functions of $\tau\in\mathbb{H}$.
For each mock modular form $h$ of weight $k$ there exists a shadow $g^*$ such that
\begin{equation}
    \hat{h}(\tau):=h(\tau)+g^*(\tau)
\end{equation}
transforms as of weight $k$ in a price of no longer holomorphic.
The shadow $g^*$ is related to a modular form $g$ of weight $2-k$
\begin{equation}
    g^*(\tau)=-(2i)^k\int_{-\bar{\tau}}^{\infty}\frac{g^c(z)}{(z+\tau)^k}\,dz
\end{equation}
where $g^c(z)=\overline{g(-\bar{z})}$.
Given $g(\tau)=\sum_{n> 0}b_nq^n$, $g^*(\tau)$ can be written as
\begin{equation}
    g^*(\tau)=\sum_{n> 0}n^{k-1}\bar{b}_n\beta_k(4n\tau_2)q^{-n}
\end{equation}
where $\tau_2=\im \tau$ and $\beta_k=\int_{t}^{\infty}u^{-k}e^{-\pi u}\,du$. Conversely, given $\hat{h}$, one determines $g$ by 
\begin{equation}
    \frac{\partial \hat{h}}{\partial \bar{\tau}}=\frac{\partial g^*}{\partial \bar{\tau}}=\tau_2^{-k}\overline{g(\tau)}.
\end{equation}
Denote the space of modular forms of weight $k$ by $M_k$, the space of completed mock modular forms by $\widehat{\mathbb{M}}_k$ and the space of weakly modular forms\footnote{
functions which transform like modular forms of weight $k$ and are holomorphic in $\mathbb{H}$, but may have singularities of type $q^{-O(1)}$ at cusps.}
by $M_k^!$, the definition induces the following maps
\begin{equation}
    0\longrightarrow M_k^!\longrightarrow \widehat{\mathbb{M}}_k \xrightarrow{\tau_2^k\frac{\partial}{\partial\bar{\tau}}} \overline{M_{2-k}}.
\end{equation}

\section{Anomaly polynomial of $T_N[S]$} \label{app:anomaly}

Consider the 6d $\cN=(2,0)$ SCFT on $T_2 \times X$ where $T^2$ is a Riemann surface and $S$ is a K\"ahler surface. 
The global symmetries include the Lorentz symmetry $SO(1,5) \subset SO(1,1) \times SU(2)_l \times U(1)_r$ and R-symmetry $SO(5) \subset SU(2)_R \times U(1)_t$. When reducing the 6d theory on $S$, one needs to perform the topological twist (MSW twist) $U(1)_{\text{tw}} = U(1)_r \times U(1)_t$ to preserve the supersymmetry. 
As studied in \cite{Chen:2022vvd}, the 2d effective theory after compactification denoted by $T_G[S]$ has $\cN=(0,4)$ supersymmetry. 
For $G=A_{N-1}$, this compactification describing $N$ M5 branes wrapping a K\"ahler 4-cycle in a Calabi-Yau threefold, which giving rise to the MSW CFT.

We will derive the anomaly polynomial of the two dimensional theory $T_G[S]$. The anomaly polynomial of $N$ M5 brane is \cite{Ohmori:2014kda} 
\begin{equation} \label{eqn:anomaly-20}
I_8[G]= N I_8(1) + (N^3-N) \frac{p_2(NW)}{24},
\end{equation}
where
\begin{equation}
I_8(1)=\frac{1}{48}\left[ p_2(NW)-p_2(TW)+\frac14\bigl(p_1(TW)-p_1(NW)\bigr)^2\right] \;,
\end{equation}
is the anomaly polynomial for a single M5-brane, $NW$ and $TW$ are the normal and tangent bundles of the worldvolume denoted by $W$, respectively. 

After twisted reduction, we expect that $SU(2)_R$ becomes the R-symmetry of $T_G[S]$ and $SU(2)_l \times U(1)_{tw} \times U(1)_t$ becomes flavor symmetries. Let $R$ and $t$ denote the $SU(2)_R$ bundle and $U(1)_t$ bundle from 6d R-symmetry, and $l$, $r$ and $T^2$ denote the $SU(2)_l$, $U(1)_r$ and $SO(1,1)$ bundle from 6d Lorentz symmetry.

The topological twist is realized by substituting $c_1(t) \to c_1(t)+c_1$, where we refer to \cite{Apruzzi:2016nfr} for more details. 
Using the fact that $p_1(t)=c_1(t)^2$ and $\int_{X}c_1^2=2\chi+3\sigma$, we perform the integral of the anomaly polynomial $I_8$ over $S$, giving
\begin{equation} \label{eq:I8M5M4}
\begin{aligned}
    I_4= &-N\frac{ (\chi+3\sigma)}{48}
   p_1(T^2)+
    \left[
   (N^3-N) \frac{2\chi+3\sigma}{6}+N\frac{(\chi+\sigma)}{4}\right]c_1(R)^2
\end{aligned}
\end{equation}
where $p_1(T^2)$ is the first Pontryagin class of the tangent bundle on $T^2$ and $c_1(R)$ is the $U(1)_R$ Cartan subalgebra of the $SU(2)_R$ R-symmetry. The central charge of $T_N[S]$ is 
\begin{align} \label{eqn:c20}
    c_R &= \frac{3}{2}(\chi+\sigma)N+(2\chi+3\sigma)(N^3-N), \nonumber \\
    c_L &= \chi N+(2\chi+3\sigma)(N^3-N), 
\end{align}

\section{Self-duality of rank two VW partition function on $\bF_2$} \label{App2}

From the character point of view, we shall prove that $h_{2,(1,0)}=h_{2,(1,1)}$ is the only case that a duality defect exists.
For the duality defect to exist, we shall have $h_{(\alpha,\beta)}=h_{(\alpha',\beta')}$ for different $(\alpha,\beta)\neq(\alpha',\beta')$.
In other words, for some $l$ and a special $(m,n)$, the VW partition functions may equal for different $(\alpha,\beta)$.
Then they shall have the same shadow. So it is enough to compare the shadow of $h$ as in Eq. \eqref{eq:thetahat} \eqref{eq:Ahat}.
From
\begin{equation}\begin{aligned}
{\vartheta^*}_{\alpha, \beta}^{m,n}+A^*_{(\alpha,\beta)}=&-\frac{1}{2}\sum_{a,b\in \bZ} E\left(((2b-\beta)n-(2a-\alpha)
  m+2(2n+(l+2)m)\frac{\im z}{\im \tau})
  \sqrt{\frac{\im \tau}{J^2_{m,n}}}\right) \\ 
&\times
w^{(l-2)(2b-\beta)+2(2a-\alpha)}\,q^{\frac{l}{4}(2b-\beta)^2+\frac{1}{2}(2b-\beta)(2a-\alpha)}
\end{aligned}
\end{equation}
we can see that for $h_{\alpha,\beta}=h_{\alpha',\beta'}$, there should exist integer $a',b'$ such that
\begin{equation}\begin{aligned}
    (2b-\beta)\frac{n}{m}-(2a-\alpha)&=(2b'-\beta')\frac{n}{m}-(2a'-\alpha')\\
    \frac{l}{4}(2b-\beta)^2+\frac{1}{2}(2b-\beta)(2a-\alpha)&=\frac{l}{4}(2b'-\beta')^2+\frac{1}{2}(2b'-\beta')(2a'-\alpha')
\end{aligned}
\end{equation}
for every integer $a,b$.
The non-trivial solution for $(a',b')$ is
\begin{equation}
\begin{pmatrix}
    a'\\b'
\end{pmatrix}=
\begin{pmatrix}
 \frac{l}{l+2 n} & -\frac{2 n (l+n)}{l+2 n} \\
 -\frac{2}{l+2 n} & -\frac{l}{l+2 n}
\end{pmatrix}\begin{pmatrix}
    a\\b
\end{pmatrix}+
\begin{pmatrix}
\frac{\alpha  (-l)+\alpha ' (l+2 n)+2 \beta  n (l+n)}{2 (l+2 n)}\\\frac{2 \alpha +\beta  l+\beta ' (l+2 n)}{2 (l+2 n)}
\end{pmatrix}
\end{equation}
in which $n$ is $n/m$.
The determinant for the matrix above is $-1$.
For $a',b'$ to be integers, $\frac{l}{l+2 n},\frac{2}{l+2 n}$ and $\frac{2 n (l+n)}{l+2 n}$ should all be integers.
$\frac{2}{l+2 n}$ being integer restricts $l+2n=\pm 1,\pm 2$.
$\frac{l}{l+2 n}$ and $\frac{2 n (l+n)}{l+2 n}$ both being integers further restricts $n=-\frac{l\pm 1}{2}$ for $l$ odd and $n=-\frac{l\pm 2}{2}$ for $l$ even.
However, $n$ should be non-negative, so the only cases left are $l=1,n=0$ and $l=2,n=0$.
In the case $l=1,n=0$, $a'$ equals $a+\frac{-\alpha+\alpha'}{2}$ and $b'=-2a-b+\alpha+\frac{\beta+\beta'}{2}$.
$\frac{-\alpha+\alpha'}{2}$ should be integer and both $\alpha$ and $\alpha'$ is valued in $\bZ_2$. So $\alpha'=\alpha$. Similarly, $\frac{\beta+\beta'}{2}$ should also be integer and so $\beta'=\beta$, i.e. this is a trivial solution and reflects a symmetry inside the summation of $h_{l=1}$.
The only case left is $l=2,n=0$. In this case
\begin{equation}
a'=a+\frac{-\alpha+\alpha'}{2},\quad
b'=-a-b+\frac{\alpha+\beta+\beta'}{2}.
\end{equation}
For $a',b'$ to be integers and $(\alpha,\beta)\neq (\alpha',\beta')$, we shall set $\alpha=\alpha'=1, (\beta,\beta')=(0,1)$.
That is to say, there is a duality $h_{l=2,(1,0)}=h_{l=2,(1,1)}$ when $n/m=0$.

\bibliographystyle{JHEP}
\bibliography{main}

\end{document}